\documentclass[12pt]{article}

\usepackage{amsmath,amssymb,amsthm,cite,enumitem,mathtools,xcolor}
\usepackage{empheq}
\newcommand*\widefbox[1]{\fbox{\hspace{2em}#1\hspace{2em}}}
\usepackage{framed}

\usepackage{cite}

\usepackage[margin=3cm]{geometry}
\usepackage{setspace}

 \usepackage[utf8]{inputenc} 
\usepackage[T1]{fontenc}
\usepackage{newtxtext} 
\usepackage{bm} 
\usepackage[normalem]{ulem}  

\usepackage{multicol}
\usepackage{authblk}

\title{Spatially flat cosmological quantum spacetimes}

\author[1]{Christian~Ga\ss}
\author[1]{Harold~C.~Steinacker}

\affil[1]{\small Faculty of Physics, University of Vienna \protect\\
Boltzmanngasse 5, A-1090 Vienna, Austria \protect\\
email: christian.gass@univie.ac.at, harold.steinacker@univie.ac.at}

\DeclarePairedDelimiter{\abs}{\lvert}{\rvert}
\DeclarePairedDelimiter{\norm}{\lVert}{\rVert}

\DeclareMathOperator{\Tr}{Tr}       
\newcommand{\lra}{\leftrightarrow}
\newcommand{\dd}{\textup{d}}
\newcommand{\ii}{\textup{i}}
\newcommand{\ee}{\textup{e}}
\newcommand{\tT}{\tilde{T}}

\newcommand{\bt}{\boldsymbol{t}}    
\newcommand{\bu}{\boldsymbol{u}} 
\newcommand{\by}{\boldsymbol{y}}    
\newcommand{\eps}{\varepsilon}      
\newcommand{\ka}{\kappa}            
\newcommand{\la}{\lambda}           
\renewcommand{\norm}[1]{\ensuremath{|\!|#1|\!|}}

\newcommand{\bC}{\mathbb{C}}        
\newcommand{\bN}{\mathbb{N}}        
\newcommand{\bR}{{\mathbb{R}}}      
       
\newcommand{\del}{\partial}         

\newcommand{\word}[1]{\quad\text{#1}\quad} 

\newcommand{\ket}[1]{\mathbin|#1\rangle}
\newcommand{\setassume}{\;|\;}
\newcommand{\gr}{>}
\newcommand{\sm}{<}

\theoremstyle{plain}
\newtheorem{thm}{Theorem}           
\numberwithin{thm}{section}         

\theoremstyle{definition}
\newtheorem{remark}[thm]{Remark}      

\def\bbbone{{\mathchoice {\rm 1\mskip-4mu l} {\rm 1\mskip-4mu l}
{\rm 1\mskip-4.5mu l} {\rm 1\mskip-5mu l}}}
\def\one{\bbbone}

\def\cL{{\mathcal L}}

\def\cQ{{\mathcal Q}}

\def\cH{{\mathcal H}}

\def\cM{{\mathcal M}}

\def\cC{{\mathcal C}}
\def\cW{{\mathcal W}}
\def\cR{{\mathcal R}}
\def\cN{{\mathcal N}}
\def\cK{{\mathcal K}}

\usepackage{graphicx}
\usepackage{floatrow}

\def\nn{\nonumber}

\def\R{{\mathbb R}} \def\C{{\mathbb C}} \def\N{{\mathbb N}}
 \def\one{\mbox{1 \kern-.59em {\rm l}}}

\def\msu{\mathfrak{su}}
\def\mso{\mathfrak{so}}
\newcommand{\End}{\mathrm{End}}
\def\hs{\mathfrak{hs}}

\makeatletter
\renewcommand{\section}{\@startsection{section}{1}{\z@}%
                       {-3.5ex \@plus -1ex \@minus -.2ex}%
                       {2.3ex \@plus.2ex}%
                       {\normalfont \large\bfseries}}
\renewcommand{\subsection}{\@startsection{subsection}{2}{\z@}%
                       {6ex \@plus -1ex \@minus .2ex}%
                       {3ex \@plus .2ex}%
                       {\centering\normalfont\normalsize\itshape}}
\renewcommand{\subsubsection}{\@startsection{subsubsection}{2}{\z@}%
                       {6ex \@plus -1ex \@minus .2ex}%
                       {3ex \@plus .2ex}%
                       {\centering\normalfont\normalsize\itshape}}                       

\usepackage{parskip}

\makeatother

\numberwithin{equation}{section}

\usepackage{hyperref}
\usepackage{xcolor}
\hypersetup{
    colorlinks,
    linkcolor={blue!30!black},
    citecolor={blue!30!black},
    urlcolor={blue!30!black}
}

\begin{document}
\maketitle 
\begin{abstract} 
\noindent 
We recently described a cosmological quantum spacetime of vanishing spatial 
curvature, which can be considered as background for the IKKT matrix model, 
assuming that the resulting gauge theory couples weakly. Building on this 
example, we construct a large class of spatially flat cosmological quantum 
spacetimes. We also elaborate on various details of their algebraic and 
semi-classical structure as well as the higher spin modes present in these 
models. In particular, we introduce the notion of approximate diffeomorphisms 
on the cosmological quantum spacetimes that stem from gauge transformations 
of the underlying matrix model, and investigate how different gauges are related 
in the semi-classical regime by approximate diffeomorphisms. Finally, we briefly 
outline how the described quantum spacetimes could be incorporated into the full 
IKKT model.
\end{abstract}
{\footnotesize
\tableofcontents}

\section{Introduction} 
It has been known for some time that the unitary irreducible  
representations of $\mso(4,2)$ in the minimal discrete 
series \cite{Fernando:2009fq}, labeled by $n\in \N_0$ and sometimes 
 called \emph{doubleton representations},
 can be used to model the 
 \emph{fuzzy hyperboloid} $H_n^4$. This is 
possible because the doubleton representations descend to irreducible 
representations of $\mso(4,1)$. In a next step, the fuzzy hyperboloid 
can be used to model covariant quantum spacetimes via certain 
projections \cite{Steinacker:2024unq,Sperling:2019xar,Sperling:2018xrm}. 
Among those quantum spacetimes, the ones that are currently best understood 
are FLRW spacetimes with negative spatial curvature ($k=-1$) and a Big 
Bounce geometry \cite{Steinacker:2017bhb, Steinacker:2024unq, 
Sperling:2019xar, Sperling:2018xrm, Manta:2025inq, Manta:2024vol, 
Battista:2022hqn}.  Such spacetimes have a spatial $SO(3,1)$-symmetry.

We recently extended the set of examples by describing a spatially flat ($k=0$) 
FLRW spacetime with a Big Bang \cite{Gass:2025tae}. In the current paper, we 
generalize this example to describe a much larger class of spatially flat FLRW 
quantum spacetimes, which have a spatial $E(3)$-symmetry.

All these cosmological quantum spacetimes should be viewed in the 
context of the IKKT matrix model \cite{Ishibashi:1996xs} formulated 
in $D=10$ dimensions. In particular, the cosmological $k=0$ quantum 
spacetime discussed in \cite{Gass:2025tae} defines 
an action 
\begin{align}
\label{eq:matrix_action_old}
 S_0[\Phi] = \Tr\big( [T^\mu, \Phi^\dagger] [T_\mu,\Phi] - m^2 \Phi^\dagger \Phi \big),
\end{align}
where $\Phi$ is a free, non-commutative scalar field on the quantum 
spacetime, and where the matrices $T^\mu$ form a static background 
that transforms under an irreducible representation of $E(3)$. $S_0[\Phi]$ 
can be seen as a toy model for the  bosonic part of the action of the 
IKKT model \eqref{eq:IKKT_S} below -- with a static instead of dynamical 
background $T^\mu$ in the regime where the matrices describing the 
extra dimensions, represented by several scalars, couple weakly.

Our paper should also be viewed as part of an ongoing program to construct 
examples of quantum spacetimes and gravity from matrix theory. We are 
aware of constructions of other cosmological spacetimes \cite{Sperling:2018xrm, 
Sperling:2019xar, Steinacker:2017bhb,Manta:2024vol,Manta:2025inq, 
MantaStein_deformed} in the framework of the IKKT model, as well as fuzzy 
de Sitter space \cite{Buric:2017yes,Gazeau:2006hj,Gazeau:2009mi}. 
For other related work on the origin of quantum spacetime from matrix theory see 
e.g. \cite{Chaney:2015ktw,Brahma:2022hjv,Brahma:2021tkh,Asano:2024def,Nishimura:2019qal,Chou:2024sgk,
Hirasawa:2023lpb}. Let us also mention recent work on the polarized IKKT model 
\cite{Hartnoll:2024csr, Komatsu:2024ydh, Martina:2025kwc, Komatsu:2024bop,Chou:2025rwy}, 
a deformation of IKKT that 
may extend the scope of consistent backgrounds.

We consider quantum spaces defined as quantized symplectic spaces $\cM$
via hermitian operators or matrices acting on a separable Hilbert space 
$\cH$. The space of "nice" operators in $\End(\cH)$ is interpreted in 
terms of quantized functions on the underlying symplectic manifold $\cM$, 
related by a quantization map 
\begin{align}
\label{eq:Q-general}
  \cQ: \quad \cC(\cM) &\to \End(\cH), \ \nn \\
  \phi(\cdot) &\mapsto \Phi.
\end{align} 
In a certain semi-classical regime, the map $\cQ$ should be invertible 
and satisfy 
\begin{align}
\cQ(\phi\psi) &\sim \cQ(\phi) \cQ(\psi), \nn \\
i\cQ(\{\phi,\psi\}) &\sim [\cQ(\phi), \cQ(\psi)], \nn \\
\word{and} 
\Tr\big(\cQ(\phi)\big) &\sim \int_{\cM} \frac{\Omega}{(2\pi)^d} \phi,  
\label{eq:Q_properties}
\end{align}
where $\sim$ indicates approximate equality in the semi-classical regime
and where $\Omega$ is the symplectic volume form on $\cM$ (which has dimension 
$2d$). Such quantization maps are not unique, but preferred maps can often be 
found by requiring their compatibility with some symmetry group, notably 
for quantized coadjoint orbits such as the fuzzy hyperboloid described in 
the following. Another way to describe quantization maps is via (quasi-)coherent 
states  $\ket{x}$ as $\Phi = \int \Omega \phi(x)|x\rangle\langle x|$; see e.g. \cite{
Steinacker:2019fcb,Steinacker:2020nva,Steinacker:2024unq} for an introduction to 
this framework.
For the specific case $\cM \cong \C P^{1,2}$ under consideration, these maps are discussed in some detail in Appendix \ref{app:k-0}.

In the case of covariant quantum spacetimes originating from the fuzzy hyperboloid, 
it turns out that the underlying symplectic space $\cM$ is the six-dimensional 
projective space ${\bC P}^{2,1}$ \cite{Sperling:2018xrm,Sperling:2019xar}, which
can be written as a local bundle 
\begin{align}
\cM  \cong \cM^{1,3} \tilde{\times} S^2,
\label{eq:Spacetime_Bundle}
\end{align}
where $\cM^{1,3}$ is a cosmological spacetime (in particular, the metric has Minkowskian 
signature) and where $\tilde{\times}$ indicates that the isometry group of the spacetime 
acts non-trivially of the local fiber. The presence of the internal two-sphere $S^2$ 
implies the existence of higher-spin degrees of freedom in the model.

Due to the non-trivial bundle structure of the symplectic space, one encounters a problem. 
Gauge transformations of the matrix model correspond to symplectomorphisms on $\cM$ 
\cite{Steinacker:2024unq} but not necessarily to diffeomorphisms on $\cM^{1,3}$. 
However, it may happen that gauge transformations are related to \emph{approximate 
diffeomorphisms} on $\cM^{1,3}$ in regimes where the contribution of the higher spin 
variables is negligible. The study of such approximate diffeomorphisms on cosmological 
$k=0$ quantum spacetimes is one of the main points of this paper. Our definitions 
and results may be extended to other covariant quantum spacetimes.

The metric of Minkowskian signature on $\cM^{1,3}$ is encoded in the \emph{matrix 
d'Alembertian} $\Box_T$ associated to the action \eqref{eq:matrix_action_old}. Its 
semi-classical version $\Box\sim\Box_T$ is in turn related to a geometric d'Alembertian 
$\Box_G$ via a \emph{dilaton} $\rho^2$, $\Box=\rho^2\Box_G$. This geometric d'Alembertian 
then yields the Minkowskian metric $G_{\mu\nu}$ on $\cM^{1,3}$, which is in general 
higher-spin valued.

We will see that on a timelike curve, the dependence of the metric tensor and 
other physical quantities on higher-spin variables can be removed by going to 
local normal coordinates. The residual higher spin components should be small 
in the semi-classical regime, more precisely in a local patch at late times 
near this timelike curve. 

The description of this semi-classical structure of cosmological $k=0$ quantum 
spacetimes forms the main body of our paper. However, we also outline how these 
spacetimes may be incorporated into the IKKT model, whose full action is given by 
\begin{align}
 S_{\rm IKKT}[T,\Psi] 
 =  \Tr\big( [T^a,T^b][T_a,T_b] + \overline{\Psi} \Gamma_a [T^a,\Psi]
 \big) . \label{eq:IKKT_S}
\end{align}
Here $a$ and $b$ run from $0$ to $9$, $T^a$ are 
hermitian matrices and $\Psi$ is a Majorana-Weyl spinor of $SO(9,1)$ with 
Grassmann-valued matrices as entries.

While the true IKKT model \eqref{eq:IKKT_S} has no quadratic "mass" term, it 
is sometimes necessary to add an $E(3)$-invariant mass term to the action 
\eqref{eq:IKKT_S} by hand to stabilize spacetime. For example, the background 
discussed in \cite{Gass:2025tae} satisfies the IKKT equations with a non-vanishing 
mass term. 

To embed the background $T^\mu$, $\mu=0,\dots,3$, describing cosmological 
quantum spacetimes into the full IKKT model, and to recover gravity 
\cite{Steinacker:2023myp}, one needs to take into account the coupling of 
$T^\mu$ to the matrices describing the extra dimensions as well as their 
self-coupling. This has been studied to some extent in the case of cosmological 
$k=-1$ quantum spacetimes \cite{Steinacker:2024unq,Battista:2023glw,MantaStein_deformed}. 
In the present paper, we will only outline how this may be extended to the case 
of $k=0$ quantum spacetimes. We expect that the classical stabilization of extra 
dimensions via $R$ charge discussed in \cite{MantaStein_deformed} will carry over 
to the $k=0$ case, while quantum effects coming from the one-loop effective action 
need to be taken into account so that the matrices defining spacetime lead to a 
physically reasonable cosmic evolution.

\subsection{Outline of the paper and summary of the main results}
Complementing several recent constructions in the case of $k=-1$ cosmological 
quantum spacetimes \cite{Steinacker:2017bhb, Sperling:2018xrm, Sperling:2019xar, 
Steinacker:2019awe, Steinacker:2024unq, Manta:2024vol, MantaStein_deformed, 
Battista:2022hqn, Battista:2023glw, Manta:2025inq}, the main aim of this paper 
is to elaborate on and generalize the example given by the action $S_0[\Phi]$ 
from \eqref{eq:matrix_action_old}, which was discussed in \cite{Gass:2025tae}.

In Section \ref{sec:so42}, we briefly recall how covariant quantum spacetimes 
arise from the fuzzy hyperboloid $H_n^4$. We work out the algebraic 
and semi-classical structure as well as the higher-spin theory on the 
cosmological quantum spacetime $\cM^{1,3}$ that arises from the local 
bundle structure $\cM=\cM^{1,3}\tilde{\times}S^2$.
We also discuss how gauge transformations can implement approximate 
diffeomorphisms on $\cM^{1,3}$ in the semi-classical regime. 

The background $T^\mu$ in \eqref{eq:matrix_action_old}, which we 
work out in Section \ref{sec:static_background} and which was partially 
investigated in the previous paper \cite{Gass:2025tae}, can be used 
as a \emph{reference background} for cosmological $k=0$ quantum 
spacetimes. We discuss its algebraic structure, which is richer in the
present case than it is for cosmological $k=-1$ quantum spacetimes: 
there are two distinct but important actions of $E(3)$, while the 
respective actions of $SO(3,1)$ in the $k=-1$ case coincide. We 
introduce appropriate derivations on $\cM^{1,3}$ that act also on 
higher-spin valued functions. In complete analogy to the $k=-1$ case, 
this derivation is used to relate vector fields on $\cM^{1,3}$ to 
tangential vector fields on $H_n^4$. 

Using the reference background $T^\mu$ as a starting point, we describe 
in Section \ref{sec:dynamical_background} dynamical backgrounds that 
respect the $E(3)$ symmetry by adding time dependent prefactors 
\begin{align}
T^0\to\tT^0:=\alpha(Y^0) T^0, 
\qquad T^i\to\tT^i:=\beta(Y^0)T^i, 
\end{align}
where $y^0\sim Y^0$ parametrizes time 
on $\cM^{1,3}$, and $\alpha,\beta$ are well-behaved functions. However, 
this description is redundant. We show that different choices $(\alpha,\beta)$ 
are related via gauge transformation generated by $\Lambda(Y^0)T^0$, corresponding 
to approximate diffeomorphisms along the time direction under mild assumptions.  
Distinguished gauges are the \emph{timelike gauge} $\beta\equiv1$, where formulas 
become particularly simple, and the \emph{covariant gauge} $\alpha=\beta$, 
in which the generators do not only transform under a representation of 
$E(3)$ but also under a representation of $SO(3,1)$.

For the dynamical backgrounds, we determine various quantities that 
are important to describe the physics in the non-commutative and the 
semi-classical regime: (i) the matrix d'Alembertian that governs the 
IKKT equations of motion, (ii) various geometric quantities that 
describe the resulting semi-classical spacetime, (iii) local normal 
coordinates, which allow to eliminate higher-spin contributions along 
a timelike curve. 

In particular, the cosmological $k=0$ spacetimes  described with the 
background $\tT^\mu$ define a FLRW geometry with cosmic scale factor 
that depends on $\alpha$ and $\beta$. We illustrate the relation between 
gauge transformation and diffeomorphisms by verifying that certain 
distinguished, gauge-equivalent backgrounds yield equivalent FLRW line 
elements in the timelike and covariant gauge. One of them is the 
background where the dilaton is constant, which is shown to yield a 
\emph{shrinking} FLRW spacetime for $k=0$, rather than an expanding 
spacetime as in the $k=-1$ case. However, we can obtain a variety of 
expanding $k=0$ FLRW spacetimes for different $\alpha$ and/or $\beta$. 
This computation also demonstrates the use and the consistency of local 
normal coordinates in the higher-spin framework \cite{Steinacker:2024unq}.

Finally, we outline in Section \ref{sec:IKKT} how the background 
$\tT^\mu$ can be incorporated in the full IKKT model. The details 
of such an incorporation are beyond the scope of the present paper 
and are left for future research. However, we concisely discuss the 
general recipe, which is based on a stabilization of dynamical 
extra-dimensions and corrections to the IKKT equations at one-loop 
and which should work similarly to the $k=-1$ case  discussed in 
\cite{Battista:2023glw, Asano:2024def, MantaStein_deformed}.

In Appendix \ref{app:k-1}, we list a number of pertinent properties 
of cosmological $k=-1$ quantum spacetimes that are needed in our 
construction of cosmological $k=0$ spacetimes. Appendix \ref{app:mmnunu} 
contains a derivation of the concrete form of the $SO(3,1)$ generators 
in various cases.

\section{Fuzzy hyperboloid and cosmological quantum spacetimes}
\label{sec:so42}
Let us 
describe how to obtain cosmological quantum 
spacetimes from matrix models. Within the scope of this paper, such 
a construction cannot be fully self-contained. We refer to the book 
\cite{Steinacker:2024unq} for a comprehensive introduction.

\subsection{Fuzzy hyperboloid} 
\label{ssc:fuzzyH}
Let $M^{ab}$, $a,b=0,\dots,5$ denote the generators of $SO(4,2)$, 
which are skew-symmetric and satisfy the commutation relations 
\begin{align}
\label{eq:commuMM}
[M^{ab},M^{cd}] &= \ii (M^{ac}\delta^{bd} - M^{ad}\delta^{bc} 
                        - M^{bc}\delta^{ad} + M^{bd}\delta^{ac}).
\end{align}
Unless specified otherwise, Latin indices $a,b,\dots$ will run from 
$0$ to $5$, Latin indices $i,j,\dots$ will run from $1$ to $3$ and 
Greek indices will run from $0$ to $3$. We use the conventions 
$(\eta^{ab})={\rm diag}(-1,1,1,1,1,-1)$ and $(\eta^{\mu\nu})={\rm diag}(-1,1,1,1)$.

In this paper, we will only consider doubleton representations $\cH_n$ of $\mso(4,2)$, 
which are labeled by $n\in\bN_0$, and assume that the $M^{ab}$ are given in one of 
these representations. There is a standard oscillator construction of the 
doubleton representations, which is described in detail in \cite{Steinacker:2024unq}.
In the doubleton representation labeled by $n$, we have the important identity 
\cite{Steinacker:2024unq}
\begin{align}
\label{eq:M_identity}
\eta_{ab} M^{ac} M^{bd} + (c\lra d) = \frac{n^2-4}{2} \eta^{cd} 
=: 2 \frac{R^2}{r^2} \eta^{cd}, \quad n=0,1,2,\dots,
\end{align}
where we introduced two dependent parameters $r^2$, $r>0$, and $R^2=r^2\frac{n^2-4}{4}$, 
which will introduce a scale on the fuzzy hyperboloid and the cosmological 
quantum spacetimes.  For further details on doubleton representations, we 
refer to the existing literature \cite{Fernando:2009fq,Steinacker:2024unq}.

The generators $M^{ab}$ can be used to introduce a set of non-commuting 
coordinate matrices 
\begin{align}
\label{eq:defX}
X^a := r M^{a5}, \quad a=0,\dots,4. 
\end{align}
Due to the identity  \eqref{eq:M_identity}, the $X^a$ satisfy the 
$SO(4,1)$-invariant constraint 
\begin{align}
\label{eq:X_SO41_constraint}
X^\mu X_\mu + {X^4}^2 = - R^2 \one,
\end{align}
which means that they can be interpreted as non-commutative coordinates 
describing the four-dimensional fuzzy hyperboloid $H_n^4$ 
of radius $R$. Strictly speaking, $R^2$ is negative for $n=0,1$ and 
vanishes for $n=2$. In the minimal case $n=0$, one can nevertheless 
describe meaningful spacetimes in some late-time regime \cite{Manta:2025inq}, 
but  to have a global semi-classical structure one should 
assume $n\gg0$  \cite{Steinacker:2024unq}. Except for a few instances where we also 
consider the minimal case $n=0$, we will always assume $n\gg0$.

\subsubsection{Semi-classical structure}
\label{sss:semiclass}
It turns out that the semi-classical structure describes six-dimensional projective 
space ${\bC P}^{2,1} := \{ z \in \bC^4\;|\; \overline{z}z =1 \}$, which can 
be viewed as coadjoint orbit of $SO(4,2)$. In other words, the fuzzy hyperboloid 
$H_n^4$ can be viewed as quantization of (the symplectic space) ${\bC P}^{2,1}$, 
which has a (local) bundle structure 
\begin{align}
\cM := {\bC P}^{2,1} \cong H^4 \tilde{\times} S^2.
\label{eq:fuzzyH_bundle}
\end{align}
Here, $H^4$ is the four-hyperboloid, $S^2$ the two-sphere and $\tilde{\times}$ 
indicates a local bundle structure and that the isometry group $SO(4,1)$ of $H^4$ 
acts non-trivially on the local fiber in a way respecting the bundle map. The 
internal two-sphere corresponds to the existence of higher-spin degrees of 
freedom in the model \cite{Steinacker:2024unq,Sperling:2019xar,Sperling:2018xrm}. 

We write $x^a \sim X^a$ and $m^{ab} \sim M^{ab}$ for the semi-classical 
versions of $X^a$ and $M^{ab}$ for $a,b=0,\dots,4$. Moreover, we have 
$\{\cdot,\cdot\} \sim -\ii [\cdot,\cdot]$. 

The algebra $\cC:=\cC(\cM)$ of functions on $\cM$ has a decomposition 
\cite{Steinacker:2024unq}
\begin{align}
\cC = \bigoplus_{s=0}^\infty  \cC^s,
\end{align}
where $\cC^s$ can be interpreted as describing the spin-$s$ degrees of 
freedom. The subspace $\cC^0$ of \emph{classical (or scalar) functions} 
consists of functions that only depend on $x^\mu$. The higher-spin 
degrees of freedom are described by two independent variables parametrizing 
the internal $S^2$.  

A useful derivation on $\cC$ is the \emph{tangential derivative}
\begin{align}
\label{tang-deriv-H4}
 \eth^a  := \frac{1}{R^2} m^{ab} \{x_b,\cdot\} 
            = - \frac{1}{R^2} x_b \{m^{ab},\cdot\}, 
 \quad a,b=0,\dots,4, 
\end{align} 
which satisfies $x_a \eth^a  =0$.
We will see later how $\eth^a$ serves to define appropriate derivations 
on the cosmological spacetimes. The trace on $\End(\cH_n)$ becomes the 
integral over $\cM$ with respect to the symplectic density as in 
\eqref{eq:Q_properties}. In order to have finite kinetic energy and inner 
products, functions $\phi\in\cC$ are subject to certain integrability 
conditions, which we tacitly assume in this work, cf.~\cite{Steinacker:2024unq} 
for details. 

Note that \eqref{eq:M_identity} implies 
\begin{align}
 \eth^a x^b = \eta^{ab} + \frac{x^ax^b}{R^2}. 
\end{align}
Moreover, for any $\phi\in\cC$, the vector field $V^a:=\{x^a,\phi\}$ is 
conserved with respect to $\eth^a$, 
\begin{align}
\eth^a V_a=0. 
\label{eq:conserved_Poisson_vf}
\end{align} 
The conservation law \eqref{eq:conserved_Poisson_vf} can be verified 
using the Jacobi identity for the Poisson bracket, the skew-symmetry of 
$m^{ab}$ and the identity \eqref{eq:M_identity}.

\subsection{Cosmological quantum spacetimes}
\label{ssc:cosmo_Q_st}
To associate a quantum spacetime to the fuzzy hyperboloid $H_n^4$, one 
needs an appropriate metric structure of Minkowskian signature. This 
structure is encoded in a \emph{matrix d'Alembertian} 
\begin{align}
\Box_T := [T^\mu,[T_\mu,\cdot]],
\label{eq:matrix_Box}
\end{align}
where $T^\mu$ is a suitable set of matrices. In the choices yielding 
cosmological spacetimes, the matrix algebra can be identified in the 
semi-classical regime with the algebra of functions on a bundle over 
spacetime as in \eqref{eq:Spacetime_Bundle}: $\cM \cong \cM^{1,3} \tilde{\times} S^2$
in analogy to \eqref{eq:fuzzyH_bundle}.
Here $\cM^{1,3}$ is the cosmological spacetime, whose coordinate functions 
are described by the semi-classical $x^\mu\sim X^\mu$, and the internal 
two-sphere $S^2$ is parametrized by $t^\mu \sim T^\mu$ with (locally) two independent 
entries. The variables $t^\mu$ are interpreted as \emph{higher spin variables}.

Specifically, the metric is constructed in the following way. For  
$\phi\in\cC(\cM)$, let $[\phi]_0 \in \cC(\cM^{1,3})$ denote its 
fiber projection. By \eqref{eq:Q_properties}, we have 
\begin{align}
 \Tr\big(Q(\phi)\big) \sim \int_{\cM^{1,3}} \Omega^{1,3} [\phi]_0 
 =: \int_{\cM^{1,3}}\rho_M [\phi]_0 \dd^4x,
\end{align}
where the volume form $\Omega^{1,3} := \rho_M \dd^4x$ on $\cM^{1,3}$ is the 
pull-back of the unique $SO(4,1)$-invariant volume form on the 4-hyperboloid 
$H^4\subset \bR^{4,1}$ spanned by the $x^a$ generators. 
It can be written explicitly as \cite{Sperling:2019xar,Gass:2025tae} 
\begin{align} 
\Omega^{1,3} 
= \frac{\dd x^0 \dots \dd x^3}{R^3 x^4} 
 = \frac{\dd y^0 \dd y^1 \dots \dd y^3}{R^3 y^0} ,
\label{eq:symplectic_form}
\end{align}
where $y^0 := x^0 + x^4$ and $y^i=x^i$.  The $SO(3,1)$-invariant quantity 
$x^4$ and the coordinates $x^\mu$ are well-adjusted to cosmological $k=-1$ 
spacetimes, while the $E(3)$-invariant quantity $y^0$ and the coordinates 
$y^\mu$ are well-adjusted to $k=0$ spacetimes.

Now consider a free, non-commutative scalar field $\Phi= \Phi(X)$ 
described by the action  $S_0[\Phi]$ from \eqref{eq:matrix_action_old}, 
which is equivalent to
\begin{align}
\label{eq:matrix_action}
 S_0[\Phi] 
 =-\Tr\big(\Phi^\dagger (\Box_T + m^2)\Phi \big).
\end{align} 
In the semi-classical regime, the action reduces to
\begin{align}
\label{eq:semicl_action}
 S_0[\Phi] \sim -\int_\cM \Omega \phi^*(\Box + m^2)\phi
  = -\int_{\cM^{1,3}} \Omega^{1,3} \big[ \phi^*(\Box + m^2)\phi \big]_0 ,
\end{align}
where the semi-classical d'Alembertian, acting on a suitable space of functions 
on $\cM$, is given by  
\begin{align}
\label{eff-dAlembertian}
\Box := -\{ t^\mu,\{t_\mu,\cdot\}\}.
\end{align}

It turns out that the right-hand side of \eqref{eq:semicl_action} 
can be given a geometric meaning when restricted to scalar 
expressions by introducing 
a \emph{dilaton} $\rho^2$ so that 
\begin{align}
 \Box=  \rho^2 \Box_G,
 \label{eq:Box_rhoBoxG}
 \end{align}
 where $\Box_G$ is a geometric d'Alembertian corresponding to a metric $G_{\mu\nu}$ 
 of Minkowskian signature, 
\begin{align}
\Box_G = - \frac{1}{\sqrt{-G}} \del_\mu \sqrt{-G} G^{\mu\nu} \del_\nu.
\end{align}

It is most convenient to describe $G^{\mu\nu}$ via a frame 
\begin{align}
 E^\alpha := \{ t^\alpha,\cdot\}, \quad 
 E^{\alpha\mu} := \{ t^\alpha,y^\mu\}, 
 \label{eq:frame_def}
\end{align}
where $y^\mu$ are the coordinates of $\cM^{1,3}$ as above.
Then, defining $\gamma^{\mu\nu} := \eta_{\alpha\beta} 
E^{\alpha\mu}E^{\beta\nu}$, one finds \cite{Steinacker:2024unq}
\begin{align}
\rho^2 = \rho_M \sqrt{|\gamma^{\mu\nu}|} \word{and}
 G_{\mu\nu} = \rho^2 \gamma_{\mu\nu}. 
 \label{eq:dilaton_and_effmetric}
\end{align}
For more general backgrounds,
the so-defined frame and metric will typically be higher-spin valued. 
Locally, one can remove their dependence on the higher-spin variables 
by changing to suitable local normal coordinates. In the present case
of (general) cosmological $k=0$ quantum spacetimes, 
the metric and frame in local normal coordinates will not depend on the 
higher-spin variables along a timelike curve, and we argue that their 
dependence on the higher-spin variables is small in the semi-classical 
regime.

\subsection{Physical scales and semi-classical regime}
\label{ssc:scales_intro}
The minimal requirement that we should impose on our model is that 
the semi-classical approximation is valid in a local patch at late times. 
In the case of cosmological $k=0$ quantum spacetimes, that means that 
for $y^0\gg0$, there must exist a sufficiently large region where 
the higher-spin degrees of freedom and other non-observed features are 
suppressed if we express physical quantities in terms of local 
normal coordinates.

Additionally, there are various physical scales that are relevant to 
determine the range of validity of the semi-classical physics in our 
matrix model description of cosmological quantum spacetimes. These are
\begin{enumerate}
 \item the effective scale of non-commutativity, 
 \begin{align}
 L_{\rm NC}^{\overline{G}} 
 = \abs{ \overline{G}_{\mu\ka} \overline{G}_{\nu\la}\theta^{\mu\nu} \theta^{\ka\la}
        }^{\frac{1}{4}},
 \label{eq:LNC}
\end{align}
 set by $\theta^{\mu\nu}=\{y^\mu,y^\nu\}$ and the classical part of the 
 effective metric $\overline{G}_{\mu\nu}$ in local normal coordinates; 
 \item the cosmic curvature scale $L_{\rm cosm} := H^{-1}$ set by the Hubble rate 
       of the FLRW spacetime associated to $\overline{G}_{\mu\nu}$ -- see 
       \eqref{eq:HR} below;
\item the order of magnitude of the higher spin components of the frame 
        and metric -- see \eqref{eq:metric_frame_hs_deviation} below. 
\end{enumerate}
For the semi-classical approximation to be valid, the cosmic curvature scale 
should be much larger than the effective scale of non-commutativity, 
\begin{align}
 L_{\rm cosm} \gg L_{\rm NC},
\end{align}
and the higher-spin components should be negligible in a reasonably large patch. 
We make these requirements precise in Subsection \ref{ssc:scales}, specifically 
in \eqref{eq:physical_growth_cond}, for concrete choices of the background.

\subsection{Gauge transformations and symplectomorphisms}
\label{ssc:symplectomorph} 
On the non-commutative level, a gauge transformation on the fuzzy 
hyperboloid $H_n^4$ is implemented by a hermitian operator 
$\Lambda\in\End(\cH_n)$. Accordingly, a real-valued function 
$\Lambda\in\cC(\cM)$ defines an infinitesimal symplectomorphism, 
i.e., an infinitesimal diffeomorphism preserving the symplectic 
form, on $\cM$ \cite{Steinacker:2024unq}, 
\begin{align}
\delta_\Lambda \phi := \{\Lambda,\phi\}, \quad \phi\in\cC(\cM).
\end{align}
This infinitesimal diffeomorphism is nothing but the Lie derivative of $\phi$ 
with respect to the Hamiltonian vector field $\xi := \{\Lambda,\cdot\}$, 
\begin{align}
 \delta_\Lambda \phi = \cL_\xi\phi .
\end{align}
Accordingly, a Hamiltonian vector field $V=\{v,\cdot\}$ on $\cM$ transforms 
under a gauge transformation $\delta_\Lambda v$ of the generator as
\begin{align}
 (\delta_\Lambda V) [\phi]  
 &= \{\{\Lambda,v\},\phi \} 
  =  \{\Lambda,\{v,\phi \}\}  - \{v,\{ \Lambda,\phi \}\} \nn\\
  &= \xi[V[\phi]] - V[\xi[\phi]] 
  = [\xi,V][\phi] \nn \\
  &= \cL_\xi V[\phi]
  \label{vectorfield-Lie}
\end{align}
for any $\phi\in\cC(\cM)$. Here, $[\xi,V]=\cL_\xi V$ is the Lie bracket of the vector 
fields $\xi$ and $V$,  which is precisely the Lie derivative of $V$ along $\xi$.
Thus, $V$ transforms covariantly (i.e., as the Lie derivative) 
under the symplectomorphism generated by $\xi$.

In the case of cosmological quantum spacetimes arising from the fuzzy 
hyperboloid, we encounter a complication: the spacetime $\cM^{1,3}$ is not 
the same as the symplectic space $\cM$, as described in Subsection 
\ref{ssc:cosmo_Q_st}. Instead, we have $\cM\sim\cM^{1,3}\tilde{\times} S^2$, 
where the internal sphere accounts for higher spin degrees of freedom 
in the model. Hence higher-spin valued quantities might not transform covariantly 
from the spacetime point of view.

The most important example is the non-covariant transformation behavior 
of higher-spin valued frames: 

Consider the frame $E^{\alpha\mu} := E^{\alpha}[y^\mu] = \{t^\alpha, y^\mu\}$ given 
by \eqref{eq:frame_def}, and $\xi^\mu := \{\Lambda,y^\mu\}=\delta_\Lambda y^\mu$. 
Assume that both of them admit decompositions 
\begin{align}
\xi^\mu=[\xi^\mu]_0+[\xi^\mu]_\hs, \quad E^{\alpha\mu}
=[E^{\alpha\mu}]_0+[E^{\alpha\mu}]_\hs 
\end{align}
into a  classical and a higher-spin component. We will show below that 
the deviation 
between the \emph{classical Lie derivative of the classical frame} 
$\cL_{[\xi]_0} [E^{\alpha\mu}]_0$ from the projection onto the spin-0 
part of the gauge variation $(\delta_\Lambda E^\alpha)[y^\mu] $ is 
given by 
\begin{empheq}[box=\widefbox]{align}
\cL_{[\xi]_0} [E^{\alpha\mu}]_0 -\big[  (\delta_\Lambda E^\alpha)[y^\mu]  \big]_0 
=  \Big[    \{ t^\alpha , [\xi^\mu]_\hs \}  - \delta_\Lambda [E^{\alpha\mu}]_\hs 
    \Big]_0.
\label{eq:hs_E_deviation}
\end{empheq}
That is to say that higher-spin components both of $\xi$ and $E$ contribute to an 
obstruction to  tensorial transformation behavior (where the rhs would vanish). 
We will assume that the classical part
 $[\xi^\mu]_0$ of $\xi^\mu$ is non-vanishing.
 
We aim to work on backgrounds (or at least in regimes) where the right-hand side 
of \eqref{eq:hs_E_deviation} is small compared to the individual terms on the left-hand 
side, in order to preserve the classical notions of differential geometry.
In particular, this is expected to hold in local normal coordinates, 
where the $\hs$ components of the frame are small.

To derive \eqref{eq:hs_E_deviation}, note first that
 \begin{align}
 [ \delta_\Lambda [E^{\alpha\mu}]_0]_0 
 &= [\xi^\rho \del_\rho [E^{\alpha\mu}]_0]_0
 = [\xi^\rho]_0 \del_\rho [E^{\alpha\mu}]_0 = \cL_{[\xi]_0} [E^{\alpha\mu}]_0 
 +  [E^{\alpha\rho}]_0  \del_\rho [\xi^\mu]_0.
 \end{align}
 Adding a zero-term in the form of the Jacobi identity, we obtain
 \begin{align}
  \cL_{[\xi]_0} [E^{\alpha\mu}]_0   
  &= \big[ \delta_\Lambda [E^{\alpha\mu}]_0 \big]_0 
  -[E^{\alpha\rho}]_0  \del_\rho [\xi^\mu]_0  \nn \\
 &\quad  - \big[ \delta_\Lambda E^{\alpha\mu} \big]_0
 + \big[ (\delta_\Lambda E^\alpha) [y^\mu]\big]_0
 + \big[ \{ t^\alpha , \xi^\mu \}\big]_0
  \nn \\
   &= \big[  (\delta_\Lambda E^\alpha) [y^\mu]\big]_0
   -\big[  \delta_\Lambda [E^{\alpha\mu}]_\hs \big]_0 
   + \big[ \{ t^\alpha , \xi^\mu \} -E^{\alpha\rho} \del_\rho  [\xi^\mu]_0) \big]_0
 \nn \\
    &= \Big[ (\delta_\Lambda E^\alpha) [y^\mu] -  \delta_\Lambda [E^{\alpha\mu}]_\hs 
    +  \{t^\alpha, [\xi^\mu]_\hs\} \Big]_0 .
 \label{eq:Lxi...}
 \end{align}

 If, on $\cM^{1,3}$, the deviation from \eqref{vectorfield-Lie} is  small in 
 a reasonable sense, then we say that the gauge transformation determined by 
 $\Lambda$ implements an \emph{approximate diffeomorphism on} $\cM^{1,3}$. 
 More precisely, we compare \eqref{eq:hs_E_deviation} to the classical Lie 
 derivative of the classical  frame $\cL_{[\xi]_0}[E^{\alpha\mu}]_0$.
 That is, 
 we use the quantity
 \begin{align}
\Delta_{\delta_\Lambda} E^{\alpha\mu} 
:= \Bigg|\frac{\cL_{[\xi]_0} [E^{\alpha\mu}]_0 -\big[  (\delta_\Lambda E^\alpha)[y^\mu]  \big]_0 
}{\cL_{[\xi]_0}[E^{\alpha\mu}]_0} \Bigg|, \qquad \cL_{[\xi]_0}[E^{\alpha\mu}]_0\neq0
\label{eq:relative_deviation_symplecto}
\end{align}
to estimate the deviation from tensoriality in the semi-classical regime.
We will elaborate this in an important example in Section \ref{ssc:approx_symplectomorph}. 

\section{Static reference background for a \texorpdfstring{\boldmath{$k=0$}}{} 
quantum spacetime}
\label{sec:static_background}

 Before describing more general cosmological $k=0$ spacetimes, let us 
first recall and elaborate on the reference background considered in 
\cite{Gass:2025tae}. Some relevant identities from the $k=-1$ case are 
collected in Appendix \ref{app:k-1}. 

The generators 
\begin{align}
T^\mu := \frac{1}{R} (M^{\mu0}+M^{\mu4}), \qquad 
Y^{\mu} := r (M^{\mu5}+\delta^{\mu0} M^{45}), 
\end{align}
where $M^{ab}$, $a,b=0,\dots,5$, are the generators of $SO(4,2)$ described 
in Section \ref{sec:so42}, were shown to yield the metric $G_{\mu\nu}$ 
of a spatially flat FLRW spacetime with a Big Bang \cite{Steinacker:2024unq,Gass:2025tae},
\begin{align} 
G_{\mu\nu} = \frac{y^0}{R} \eta_{\mu\nu}.
\end{align}
Here, $y^0>0$ is the conformal time. Using the cosmological time $t:= \tfrac{2}{3\sqrt{R}}
 {y^0}^\frac{3}{2}$, the metric $G_{\mu\nu}$ corresponds to the line element 
\begin{align} 
\dd s_G^2 = -\dd t^2 + a(t)^2 \dd\by^2 
 \ \word{with} \ a(t)= \big( \tfrac{3t}{2R}  \big)^\frac{1}{3}.
\end{align}

The $T^\mu$ generators will define a matrix background\footnote{These are special 
cases of the "pure backgrounds" considered in \cite{Steinacker:2024unq}.} in the IKKT model, 
while the $Y^\mu$ generators are considered as (quantized) local coordinate functions 
on spacetime. The matrices $T^\mu$ and $Y^\mu$ satisfy the commutation relations\footnote{These 
commutation relations are very similar to those of $k=-1$ quantum spacetime upon zooming in to 
a locally flat patch  \cite{Steinacker:2026qzk}.} 
\begin{align} \label{eq:commut_rel_TY}
[T^i,T^j] &= 0 , \quad  [T^0,T^i] = -\frac \ii R T^i, \quad
[T^\mu,Y^\nu] = \ii \frac{Y^0}{R} \eta^{\mu\nu}
\end{align}
as well as 
\begin{align}\label{eq:commut_rel_YY}
[Y^0,Y^i] = \ii r^2 R T^i, \quad 
[Y^i,Y^j] 
= -\ii r^2 M^{ij}.
\end{align}

The identity \eqref{eq:M_identity} implies the 
following constraints on the $T^\mu$ and $Y^\mu$ generators:
\begin{align} 
T^i T_i - \frac{1}{r^2R^2} {Y^0}^2 &= 0, \quad
T^\mu Y_\mu  + Y^\mu T_\mu =0.
\label{eq:TY_constraints}
\end{align}

\subsection{\texorpdfstring{$E(3)$}{}-subalgebra, invariant quantities and Casimir 
operators}
Clearly, the set $\{ M^{ij}, T^i \;|\; i,j=1,2,3\}$ 
can be interpreted as generators of an $E(3)$-subalgebra
because
\begin{align}
[M^{ij},T^k] 
= \ii\big( \delta^{ik} T^j -\delta^{jk} T^i\big). 
\label{eq:commu_MT}
\end{align}
They act via commutators on the space of matrices $\End(\cH)$.
A priori, it is clear that all quantities invariant under 
$SO(3)\subset E(3)$ must be functions of 
\begin{align}\label{eq:SO3_inv_list}
Y^0,\; T^0,\; Y^k Y_k,\; T^k T_k, \;Y^k T_k \word{and} T^kY_k.
\end{align} 
However, there are several redundancies due to the constraints 
\eqref{eq:commut_rel_TY} and the identity 
\begin{align} 
\frac{4\ii}{R} Y^0 = [T^\mu,Y_\mu] = 2 T^\mu Y_\mu.
\end{align}
Eliminating the redundancies, we are left 
with three independent linear or quadratic 
$SO(3)$-invariant quantities, in terms of which all $SO(3)$-invariant 
quantities can be expressed:
\begin{align}
Y^0, \; T^0 \word{and} Y^kY_k.
\end{align}

Among these three, only $Y^0$ is invariant under $E(3)$. To see this, 
note that by \eqref{eq:commut_rel_TY}, $Y^0$ commutes with the generators 
$T^i$ of the translations, while $T^0$ and $Y^kY_k$ don't. Therefore, any 
$E(3)$-invariant quantity must be a function of $Y^0$ only.

Note that both $T^i$ and $T^\mu$ transform as vectors under 
$E(3)$. That is, both of them carry a representation of $E(3)$. In the 
four-dimensional representation on the space with basis $T^\mu$, the
translations are implemented by a function $c_k T^k$, so that 
\begin{align}
\begin{rcases*}
\delta T^0 = \ii [c_k T^k,T^0]= -\frac{1}{R} c_k T^k \\
\delta T^i = \ii [c_k T^k, T^i]=0  
\end{rcases*}
\quad\Rightarrow\quad   \delta T^\mu = \Lambda^\mu_{\;\nu} T^\nu.
 \label{eq:covariance}
\end{align}

In terms of the generators of $E(3)$, the two quadratic Casimir 
operators of $E(3)$ are given by 
\begin{align}
C_1 = T^i T_i \word{and} C_2 = \eps_{ijk} T^i M^{jk}.
\label{eq:E3_Casimir}
\end{align}
Their values on the $SO(4,2)$-doubleton representation labeled by $n$ are 
\begin{align}
 C_1 &= \frac{{Y^0}^2}{r^2R^2}   \word{and}
 C_2 =  -\frac{n}{rR} Y^0 .
 \label{eq:Casimir_values}
\end{align}
 The first Casimir simply follows from the normalization of the $T^i$, while 
 the second follows from the identity \eqref{eq:epsMM}.\footnote{Note that to derive 
 $C_2$ from \eqref{eq:epsMM}, one needs to keep track of signs carefully 
 because the Minkowski metrics relating $X^0$ to $X_0=-X^0$ and $Y^0$ to 
 $Y_0=-Y^0$ do not coincide.}
 
\begin{remark}
 We want to point out an interesting coincidence. Observe that a 
 Yang-Mills type action of the form
\begin{align}
S[T] :=&\; \Tr\big( [T^\mu,T^\nu] [T_\mu,T_\nu] \big) 
\end{align}
is manifestly invariant under the global $E(3)$ symmetry 
\begin{align}
S[T^0,T^i+c^i\one]=S[T^0,T^i] \word{and} 
S[T^0, R^{i}_{\;j} T^j]=S[T^0,T^i], 
\end{align}
where $c^i$ are constants and where $R^i_{\;j}$ is a rotation. 
This transformation is distinct from the above gauge invariance under $E(3)$, 
in contrast to the $SO(3,1)$  symmetry of the $k=-1$ case. The transformation 
\begin{align}
T^i\to T^i+\delta T^i := T^i + c^i \one = T^i + \ii [\Lambda_{\boldsymbol{c}},T^i]
\end{align}
for $k=0$ is also equivalent to a gauge transformation, generated by $\Lambda_{\boldsymbol{c}}= \frac{R }{Y^0} c_i Y^i$.
Therefore this symmetry does not lead to physical zero modes.

\end{remark}

\subsection{Explicit form of the $SO(3,1)$ generators}
\label{ssc:so31}
It was shown in \cite{Gass:2025tae} that in the minimal representation $n=0$, the 
generators of rotations $M^{ij}$ are given by\footnote{ Note that the generators $T^\mu$ 
are no longer hermitian for $n=0$ because $R=\ii r$ in this case. The $SO(3,1)$ generators, 
however, remain hermitian. $T^\mu$ can be made hermitian for $n=0$ via the redefinition 
$T^\mu = \frac{1}{r}(M^{\mu0}+M^{\mu4})$, and adjusting the formulas for the $SO(3,1)$ 
generators accordingly.} 
\begin{align}
M^{ij} = \frac{R}{Y^0} \big( T^i Y^j - T^j Y^i\big), \quad n=0.
\label{eq:minimal_M}
\end{align}
In fact, this form can only be correct in the case $n=0$ because 
$M^{ij}$ from \eqref{eq:minimal_M} satisfies $C_2=0$, where $C_2$ 
is the second Casimir operator from \eqref{eq:E3_Casimir} or 
\eqref{eq:Casimir_values}, respectively.

Determining $M^{ij}$ explicitly for generic $n$ is beyond the scope of this paper. 
It is clear, that $M^{ij}$ must be of the form 
\begin{align}
 M^{ij} = c (T^i Y^j - T^j Y^i) + d \eps^{ijk} Y_k + f \eps^{ijk} T_k,
\end{align}
where $c$, $d$ and $f$ are functions of the $SO(3)$-invariant quantities 
$Y^0$, $T^0$ and $Y^kY_k$.
However, their semi-classical limits $m^{ij}\sim M^{ij}$ for $n\gg0$ 
and also in the minimal case $n=0$ can be determined explicitly. In 
Appendix \ref{app:mmnunu}, we derive 
\begin{empheq}[box=\widefbox]{align}
m^{i0} &=  \frac{R}{y^0} \times
\begin{cases} 
   \tfrac{{y^0}^2+\by^2+R^2}{2y^0} t^i- t^0 y^i , &\quad n=0;  \\
\tfrac{{y^0}^2+\by^2-R^2}{2{y^0}} t^i - t^0 y^i 
+ \frac{R}{{y^0}} \eps^{ijk} y_j t_k  , &\quad n\gg0;
\end{cases}
\nn \\
 m^{ij} 
 &= \frac{R}{y^0} \times \begin{cases} 
  t^i y^j - t^j y^i, &\hspace{62pt} n=0; \\
   t^i y^j - t^j y^i + R \eps^{ijk} t_k , &\hspace{62pt}n\gg0. 
 \end{cases}
 \label{eq:m_gens}
\end{empheq}
A simple explicit realization of the minimal $n=0$ representation in terms of 
three canonical generators is given in \cite{Ho:2025htr}.

\subsection{Fiber projection}
\label{ssc:fiber_proj}
We would like to average higher-spin valued quantities over the 
internal $S^2$. As introduced in Subsection \ref{ssc:cosmo_Q_st}, 
we denote this average by $[\cdot]_0$. The starting point is to 
average linear quadratic expressions in the $t^\mu$ variables.

Using the semi-classical version of the identity \eqref{eq:TY_constraints}, 
 we parametrize $\bt = \tfrac{y^0}{rR} \bu$, with $\bu^2=1$, i.e., 
 $\bu\in S^2\hookrightarrow \bR^3$, and $t^0 = \tfrac{1}{rR} (\by\bu)$. 
 Therefore, $[t^\mu]_0 = 0$ and, using the identity 
\begin{align}
[u^iu^j]_0 = \frac{1}{4\pi} \int_{S^2} \dd\bu u^iu^j  =\tfrac{1}{3} \delta^{ij},
\end{align}
we find 
\begin{empheq}[box=\widefbox]{align}
&[t^\mu t^\nu]_0 
= \frac{1}{3r^2R^2}\kappa^{\mu\nu}
\label{eq:S2_avg_tt}
\end{empheq}
with $\kappa^{00} = \by^2$, $\kappa^{i0}=\kappa^{0i} = y^0y^i$ and 
$\kappa^{ij} = {y^0}^2\delta^{ij}$.

Note that $\kappa^{\mu\nu}$ is transversal in the sense $\kappa^{\mu\nu} y_\nu=0$.
As a consequence of \eqref{eq:S2_avg_tt} and \eqref{eq:m_gens}, the various 
components of the average $[t^\alpha m^{\mu\nu}]_0$ are given by 
\begin{align}
[t^0 m^{i0}]_0 
&=  \frac{1}{3r^2R y^0} \times
\begin{cases} 
   -\tfrac{y^\mu y_\mu-R^2}{2} y^i ;  \\
 -\tfrac{y^\mu y_\mu +R^2}{2}   y^i;
\end{cases}
\nn \\
[t^l m^{i0}]_0 &=  \frac{1}{3r^2R } \times
\begin{cases} 
   \tfrac{{y^0}^2+\by^2+R^2}{2} \delta^{l i}-  y^l y^i;  \\
\tfrac{{y^0}^2+\by^2-R^2}{2}  \delta^{li} -  y^l y^i - R\eps^{ilj}  y_j;
\end{cases}
\nn \\
 [t^0 m^{ij}]_0 
 &= \frac{1}{3r^2R} \times \begin{cases} 
  0; \\
   R \eps^{ijk} y_k;
 \end{cases}
\nn \\
 [t^l m^{ij}]_0 
 &= \frac{y^0}{3r^2R} \times \begin{cases} 
   \delta^{li} y^j - \delta^{lj} y^i; \\
   \delta^{li} y^j - \delta^{lj} y^i + R \eps^{ijl},
 \end{cases} 
 \label{eq:m_gens_fiber_proj}
\end{align}
where in all four equations, the upper case corresponds to the minimal 
representation $n=0$ and the lower case is valid for $n\gg0$.

Fiber projections of polynomials of higher order in $t^\mu$ can be 
computed using Wick type formulas as in \cite{Steinacker:2024unq,Steinacker:2019awe}.

\subsection{Derivatives}
\label{ssc:derivatives}
In this subsection, we will construct suitable derivative operators acting on 
the algebra of $\hs$-valued functions on spacetime, analogous to the construction 
for $k=-1$ in \cite{Steinacker:2024unq}. We will always assume $n\gg0$.  To 
obtain more transparent formulas, it is sometimes convenient to use the variables 
$x^a$ suitable to describe the fuzzy hyperboloid and $k=-1$ spacetimes instead of 
$y^\mu$. However, in all formulas of the following considerations, $x^a$ should 
be understood as $x^a(y^\mu)$ via the relations 
\begin{align}
x^0 = \frac{{y^0}^2+ \by^2 + R^2}{2 y^0}, \quad \boldsymbol{x}=\by 
\word{and}
x^4 = \frac{{y^0}^2- \by^2 - R^2}{2 y^0}.
\label{eq:xy_rel}
\end{align}

In terms of the variables 
$y^\mu$ and $x^4$, the tangential derivatives \eqref{tang-deriv-H4} on $H^4$ become 
\begin{empheq}[box=\widefbox]{align}
\eth^0 &= -\frac{1}{R^2} \Big( m^{j0} \{y_j,\cdot\} - R t^0 \{ x^4,\cdot\} \Big); \nn \\
\eth^i &= -\frac{1}{R^2} \Big( m^{i0} \{y^0,\cdot\} - m^{ij} \{ y_j,\cdot\} 
        - Rt^i \{x^4,\cdot\} \Big); \nn \\
\eth^4 &= -\frac{1}{R^2} \Big( -Rt^0 \{y^0,\cdot\} + (Rt^i-m^{i0})\{y_i,\cdot\} 
                        + Rt^0 \{x^4,\cdot\}\Big), 
\label{eq:eth}
\end{empheq}
where we may insert $m^{ij}$ and $m^{i0}$ for $n\gg0$ from \eqref{eq:m_gens}.
The action of $\eth^a$ on $y^\mu$ is given by 
\begin{align}
\eth^0 y^0 
&= - 1 + \frac{x^0y^0}{R^2}, \qquad 
\eth^i y^0 
= \frac{y^0y^i}{R^2}, \qquad
\eth^4 y^0  
= 1+ \frac{x^4{y^0}}{R^2}, 
\nn \\ \word{and}
\eth^a y^j &= \eta^{aj} + \frac{x^a y^j}{R^2}.
\label{eq:eth_y0}
\end{align}

Moreover, Eq.~\eqref{eq:eth} and the Poisson brackets \eqref{eq:Poisson_x4} 
yield 
\begin{align}
\eth^0 t^0 
 &= \frac{x^0t^0}{R^2}  ; \nn \\ 
\eth^0 t^j 
&= \frac{1}{R^2} \Big( x^0 t^j + \frac{R}{y^0} \big( - Rt^j + \eps^{jkl} y_k t_l \big)   \Big); \nn \\
\eth^i t^0 
&= -\frac{1}{R^2} \Big( -t^0 y^i + \frac{R}{y^0} \big( - R t^i + \eps^{ikl} y_k t_l \big)\Big); 
\nn \\
\eth^i t^j 
&=-\frac{1}{R^2} \Big( - y^i t^j + R \eps^{ijk} t_k  \Big); \nn\\
\eth^4 t^0 
&= \frac{x^4 t^0}{R^2}; \nn \\
\eth^4 t^j 
&= -\frac{1}{R^2} \Big( -x^4 t^j + \frac{R}{y^0} \big( -R t^j + \eps^{jkl} y_k t_l \big) \Big).
\end{align}

A derivative $\del^\mu$ on $\cM^{1,3}$ that is appropriately applicable to 
$\hs$-valued functions should satisfy 
\begin{align}
\del^\mu y^\nu = \eta^{\mu\nu}.
\label{eq:dely_eta}
\end{align}
It is easy to see that 
\begin{empheq}[box=\widefbox]{align} 
 \del^\mu :=&\; \eth^\mu - \frac{x^\mu}{y^0} (\eth^0+\eth^4) 
  = \eth^\mu - \frac{y^\mu-\eta^{\mu0} x^4}{y^0} (\eth^0+\eth^4) 
 \label{eq:hsdel}
\end{empheq}
fulfills this condition. Note that $\del^0 = \frac{1}{y^0} (x^4\eth^0 - x^0 \eth^4)$. 
Applied to the higher-spin variables $t^\mu$, we obtain
\begin{align}
\del^0 t^0 = 0, \quad 
\del^0 t^j = \frac{1}{y^0} \big(- t^j +\frac{1}{R} \eps^{jkl} y_k t_l \big) = - \del^j t^0
\word{and}
\del^i t^j = -\frac{1}{R} \eps^{ijk} t_k. 
\label{eq:del_t_ngg0}
\end{align}
We thus have $\del^\mu t^\nu + \del^\nu t^\mu=0$.

\begin{remark}
 Note that \eqref{eq:eth_y0} is only true for $n\gg0$. Inserting 
 instead $m^{ij}$ and $m^{i0}$ for $n=0$ given in \eqref{eq:m_gens} yields 
 $\eth^0 y^0= \frac{x^0y^0}{R^2}$ and $\eth^4 y^0 = \frac{x^4y^0}{R^2}$. 
 Hence there is no straightforward analog of \eqref{eq:hsdel} that satisfies 
 \eqref{eq:dely_eta} for $n=0$. The problem arises because the minimal 
 representation $n=0$ admits a semi-classical structure only
 in the limit $y^0\gg0$, so $\pm 1$ is subleading to $\frac{x^a y^0}{R^2}$ and we 
 need to make a suitable approximation for the generators $m^{\mu\nu}$ to obtain 
 reasonable tangential derivatives. This is beyond the scope of the present paper.
\end{remark}

\subsection{Reconstruction of divergence-free vector fields}
In this subsection, we will again assume $n\gg0$. In the case of a cosmological 
$k=-1$ spacetime, it has been shown that the pushforward $V^\mu$ of a tangential
vector field $V^a$ on $H^4$ to the spacetime $\cM^{1,3}_{k=-1}$ satisfies 
\begin{align}
 \eth_a V^a 
 &= \sinh(\eta) \del_\mu (\rho_{M,k=-1} V^\mu), 
 \label{eq:VV_corresp_-1}
\end{align}
where $\eth_a$ is the tangential derivative on $H^4$ defined in \eqref{tang-deriv-H4}; 
and that conversely any vector field $V^\mu$ on $\cM^{1,3}_{k=-1}$ possesses 
a lift to $H^4$ so that \eqref{eq:VV_corresp_-1} is satisfied \cite{Steinacker:2024unq}. 
Note that the vector fields may be $\hs$ valued.
Therefore, there is a correspondence between tangential and divergence-free 
vector fields on $H^4$ to divergence-free vector fields on $\cM^{1,3}_{k=-1}$.
In the following, we establish the analogous correspondence for the $k=0$ case.

To do so, suppose that $V^\mu$ is a (possibly $\hs$-valued) vector field on $\cM^{1,3}$. 
We can lift $V^\mu$ to a vector field $\tilde{V}^a$ on $H^4$ by defining: 
\begin{align}
\tilde{V}^0 := V^0 + \frac{x_\mu V^\mu}{y^0}, \quad 
\tilde{V}^i := V^i,\;i=1,2,3,\word{and}
\tilde{V}^4 := - \frac{x_\mu V^\mu}{y^0}.
\label{eq:vf_lift}
\end{align}
The so-defined $\tilde{V}^a$ is tangential on $H^4$: 
\begin{align}
x_a \tilde{V}^a &= x_\mu V^\mu + \frac{\eta^{00} x^0-\eta^{44}x^4}{y^0} x_\mu V^\mu =0.
\end{align}
Moreover, we have 
\begin{empheq}[box=\widefbox]{align}
 \eth_a \tilde{V}^a = y^0\del_\mu \Big( \frac{1}{y^0} V^\mu\Big) 
 =  R^3 y^0\del_\mu \big( \rho_M V^\mu\big)
 \label{eq:vf_div_rel}
\end{empheq}
with $\del_\mu=\eta_{\mu\nu}\del^\nu$ as in \eqref{eq:hsdel}. The relation 
\eqref{eq:vf_div_rel} can be proved by a short computation, which uses 
\eqref{eq:M_identity}:
\begin{align}
 y^0\del_\mu \Big( \frac{1}{y^0} V^\mu\Big) 
 &= \del_\mu V^\mu - \frac{V^0}{y^0} \nn \\
 &= \eth_\mu V^\mu -\frac{1}{y^0} \Big( V^0+ x_\mu(\eth^0+\eth^4) V^\mu\Big) \nn \\
 &= \eth_\mu V^\mu -\frac{1}{y^0} \Big( V^0+  (\eth^0+\eth^4) x_\mu V^\mu 
  - V_\mu (\eth^0+\eth^4) x^\mu \Big) \nn \\
 &= \eth_\mu V^\mu -\frac{1}{y^0} \Big(   (\eth^0+\eth^4) x_\mu V^\mu 
  -  \frac{y^0}{R^2} x_\mu V^\mu  \big) \Big) \nn \\  
 &= \eth_a \tilde{V}^a + x_\mu V^\mu\Big(   (\eth^0+\eth^4) \frac{1}{y^0}
  +  \frac{1}{R^2} \big) \Big) \nn \\    
  &= \eth_a \tilde{V}^a.
\end{align}

Next, consider a -- possibly $\hs$-valued -- function $\phi=\phi(y,t)\in\cC$, which defines
a vector field $V^\mu := \{y^\mu, \phi\}$. The so-defined $V^\mu$ is 
conserved in the sense 
\begin{empheq}[box=\widefbox]{align}
\del_\mu \big(\rho_M V^\mu\big)=0, \quad V^\mu = \{y^\mu,\phi\}.
\label{eq:conserved_vf}
\end{empheq} 
This follows as a corollary from 
\eqref{eq:vf_div_rel}. Instead of computing $\del_\mu \big(\rho_M V^\mu\big)$, 
we may compute $\eth_a \tilde{V}^a$. Then, expressing $y^0$ in terms of 
$x^0$ and $x^4$, a short computation yields 
\begin{align}
 \eth_a \tilde{V}^a = - \frac{m^{0b}+ m^{4b}}{R^2} 
 \big\lbrace x_b, \tfrac{1}{2y^0} \{x^a x_a,\phi\}\big\rbrace =0. 
\end{align}

For $\phi=y^\nu$ and $\theta^{\mu\nu} = \{y^\mu,y^\nu\}$, \eqref{eq:conserved_vf} 
implies in particular
\begin{align}
\del_\mu (\rho_M \theta^{\mu\nu}) &= 0.
\end{align} 

\section{Dynamical background and general \texorpdfstring{\boldmath{$k=0$}}{}  
quantum spacetime}
\label{sec:dynamical_background}
We now describe more general cosmological $k=0$ quantum spacetimes that 
arise from the fuzzy hyperboloid. To do so, we start by considering  
more general matrix backgrounds respecting the $E(3)$ symmetry.

\subsection{Generalized \texorpdfstring{$E(3)$}{}-covariant background}

The generalized translation generators $\tT^i$ of $E(3)$ should commute with each other and
with the $E(3)$-invariant quantity $Y^0$; and additionally transform as 
vector under the $SO(3)$-subgroup generated by the $M^{ij}$. Any such vector 
$\tT^i$ must be a linear combination of $T^i$, $Y^i$ and $\eps^{ikl} Y_k T_l$ 
with coefficients that may depend on $Y^0$. That is, we make the ansatz 
\begin{align}
\tT^i &:= \beta(Y^0) T^i  + \gamma(Y^0) Y^i + \delta(Y^0) \eps^{ikl} Y_k T_l.
\end{align} 
Using the commutation relations \eqref{eq:commut_rel_TY} and \eqref{eq:commut_rel_YY},
one readily verifies that 
\begin{align}
[\tT^i,Y^0]=[\tT^i,\tT^j] =0
\end{align}
together imply $\gamma=\delta=0$. 

We will see below that the function $\beta$ is sufficient to describe 
a generic cosmological $k=0$ spacetime -- as should be, because the 
latter is entirely determined by the time-dependent scale factor. 
However, it is convenient to consider as well a deformed generator 
$\tT^0 := \alpha(Y^0) T^0$. 

Summing up, we define a new, dynamical (i.e., $Y^0$-dependent) background 
$\tT^\mu$ via 
\begin{empheq}[box=\widefbox]{align}
\tT^0= \alpha(Y^0) T^0; \quad \tT^i = \beta(Y^0) T^i.
\label{eq:new_generators.}
\end{empheq}

In fact, we will show in Subsection \ref{ssc:approx_symplectomorph} that there 
are gauge transformations -- defining approximate diffeomorphisms on $\cM^{1,3}$ -- 
which relate well-behaved gauge configurations $(\alpha,\beta)$ to other 
well-behaved gauge configurations $(\tilde{\alpha},\tilde{\beta})$. Here, 
"well-behaved" essentially means that $\alpha$ and $\beta$ should, at late times, 
satisfy some monotonicity conditions and not behave too differently. 
Thus, the semi-classical structures for two gauge configurations
related by such an approximate diffeomorphism are equivalent. 

The choices $\alpha\equiv1$ or $\beta\equiv1$ yield 
simple expressions for many interesting objects like the frame, 
the effective metric or the local normal coordinates. On the 
other hand, one might say that the choice $\alpha\equiv\beta$ 
is preferred because then, as in the reference case, the 
generators additionally carry a representation of $SO(3,1)$.

We call $\beta\equiv1$ the \emph{timelike gauge}, 
$\alpha\equiv1$ the \emph{spacelike gauge} and $\alpha\equiv\beta$
the \emph{covariant gauge}.

\subsection{Gauge symmetries}
\label{ssc:gauge}

An important problem is to understand how diffeomorphisms of spacetime arise from gauge transformations. 
The present setting suggests to focus on diffeos which respect $E(3)$. These arise from 
 gauge transformations generated by $\Lambda(Y^0)T^0$, which preserve the structure of the 
background due to the commutation relations \eqref{eq:commut_rel_TY}:
\begin{align}
 \ee^{-\ii\Lambda(Y^0)T^0} \alpha(Y^0) T^0  \, \ee^{\ii\Lambda(Y^0)T^0} 
 &= \tilde{\alpha}(Y^0) T^0; \nn \\
  \ee^{-\ii\Lambda(Y^0)T^0} \beta(Y^0) T^i \, \ee^{\ii\Lambda(Y^0)T^0} 
 &= \tilde{\beta}(Y^0) T^i. 
 \label{eq:gauge_def}
\end{align}
 
 Since $[\Lambda T^0, [\Lambda T^0,\tT^\mu]] \neq0$
for a generic function $\Lambda$, it is not easy to determine 
the full action of the gauge transformation on the generators $\tT^\mu$. However, 
it is sufficient to determine the infinitesimal action 
 \begin{align}
 \delta_{\Lambda T^0} \tT^\mu 
 &= \ii [\tT^\mu, \Lambda(Y^0) T^0].
 \end{align}
In detail, we have 
  \begin{align}
 \delta_{\Lambda T^0} \tT^0
 &= \frac{Y^0 W(\alpha,\Lambda) }{R}  T^0 \word{and}
  \delta_{\Lambda T^0} \tT^i
 =  -\frac{\Lambda (Y^0\beta)'}{R }  T^i, 
 \label{eq:delta_LaT0}
 \end{align} 
where $W(f,g) := fg'-f'g$ is the Wronskian and where the prime indicates 
a derivative with respect to $Y^0$.

We note the interesting 
fact that for non-trivial $\Lambda$, we have $\delta_{\Lambda T^0}\tT^i=0$
 if and only if $\beta =\tfrac{const.}{Y^0}$. The semi-classical structure 
 corresponding to this particular choice of $\beta$ is therefore not 
 related to the structures for other $\beta$ by an approximate diffeomorphism. 
 In fact, we will see that $\beta=\tfrac{const.}{Y^0}$ corresponds to a 
 degenerate frame and metric.
 
In all other cases, the infinitesimal gauge transformation is precisely 
the one relating backgrounds described by $(\alpha,\beta)$ to backgrounds 
described by $(\tilde{\alpha},\tilde{\beta})$ for appropriate $\tilde{\alpha},
\tilde{\beta}$.

\subsection{Commutation relations and matrix d'Alembertian}
The new generators \eqref{eq:new_generators.} satisfy the commutation 
relations
\begin{align}
[\tT^i,\tT^j] &= 0; \quad
&&[\tT^0,\tT^i] = -\frac{\ii \alpha }{R} \big( Y^0 \beta \big)' T^i;
\nn \\
[\tT^0, Y^0] &= -\ii \frac{Y^0 \alpha }{R};
&&[\tT^0,Y^j] = \ii r^2 R \alpha' T^j T^0; \nn \\
[\tT^i,Y^0] &= 0; 
&&[\tT^i,Y^j] 
          = \ii \frac{Y^0}{R}\beta \delta^{ij} + \ii r^2 R \beta' T^i T^j, 
\end{align} 
and the double-commutators are given by (no summation)
\begin{align}
[\tT^0,[\tT^0,\tT^j]] 
&= -\frac{\alpha}{R^2} \Big(  Y^0 \alpha \; \big( Y^0 \beta \big)'\Big)' T^j
\nn \\
[\tT^i,[\tT^i,\tT^j]] &= 
[\tT^i,[\tT^i,\tT^0]] = [\tT^0,[\tT^0,\tT^0]]= 0.
\label{eq:double_commu_U}
\end{align}

Thus, applying the matrix d'Alembertian $\Box_{\tT} := [\tT^\mu,[\tT_\mu,\cdot]]$ on 
the background yields 
\begin{empheq}[box=\widefbox]{align}
 \Box_{\tT} \tT^0 =0 \word{and}
 \Box_{\tT} \tT^j = \frac{\alpha}{R^2 \beta} \big(  Y^0 \alpha \; ( Y^0 \beta )'\big)'   \tT^j .
\label{eq:BoxU_U}
\end{empheq}

The specific generators $U^\mu$ obtained by setting $\alpha=\beta=\frac{rR}{Y^0}$ 
yield a degenerate frame and metric, and we will not consider them in the following 
as physical background. However, they are very convenient to parametrize the internal 
$S^2$ because they are normalized, $U^i U_i = 1$ (see also Subsection \ref{ssc:fiber_proj}). 
Therefore, if $\hs$-valued quantities are expressed in terms of $u^\mu\sim U^\mu$, 
it is easy to estimate the contribution of the $\hs$ part. We will thus 
use $u^\mu$ instead of $t^\mu$ to expand the $\hs$ components of physical quantities.

We observe that the reference background $T^\mu$ corresponding to 
$\alpha\equiv\beta\equiv 1$ satisfies the equations $\Box_T T^0=0$ 
and $\Box_T T^i = \frac{1}{R^2} T^i$. It is therefore a classical solution of 
the IKKT model with mass terms for the spatial generators $T^i$.

\subsection{Higher-spin valued frame, dilaton and effective metric} 
The frame arising from the general background $\tilde T^\mu \sim \tilde t^\mu$ reads 
\begin{align}
(E^{\alpha\mu}) :=  \{\tilde{t}^\alpha, y^\mu\}
&= \frac{y^0}{R} \begin{bmatrix}
-\alpha & \alpha' (\by\bu) \bu^t 
\vspace{12pt}\\
\boldsymbol{0} & \beta \one + y^0\beta' \bu\bu^t
\end{bmatrix}
\label{eq:frame_hs}
\end{align} 
with $u^\mu=\tfrac{rR}{y^0} t^\mu$ so that $\bu^2=1$ and 
$u^0=\frac{(\bu\by)}{y^0}$. Note that the inverse 
of a rank one perturbation of the identity is 
\begin{align}
(\one+ a \boldsymbol{v}\boldsymbol{v}^t)^{-1} 
=\one- \frac{a \boldsymbol{v}\boldsymbol{v}^t }{1+a \boldsymbol{v}^2}  , 
\end{align}
so that the inverse  of the frame is formally given by 
\begin{align}
(E_{\alpha\mu}) 
&= \frac{R}{y^0} \begin{bmatrix}
 -\frac{1}{\alpha}
 &\frac{\alpha' }{ \alpha (y^0\beta)'} (\by\bu) \bu^t\vspace{12pt}\\
 \boldsymbol{0} 
 & \frac{1}{\beta} \Big( \one - \frac{y^0\beta' }{(y^0\beta)'} 
 \bu\bu^t \Big)
\end{bmatrix}.
\end{align} 
We now see explicitly 
that the frame is  
non-degenerate if $(y^0\beta)'\neq0$, hence if $\beta\neq\tfrac{const.}{y^0}$.
In the following, we will exclude this choice of $\beta$, although we will 
sometimes comment on it. 
The determinant of the frame and the dilaton can easily be determined using 
\eqref{eq:dilaton_and_effmetric}, 
\begin{align}
\det(E^{\alpha\mu}) 
&= -\frac{{y^0}^4\alpha \beta^2 (y^0\beta)'}{R^4}
\word{and} 
\rho^2  =  \frac{{y^0}^3\alpha \beta^2 (y^0\beta)'}{R^3}.
\label{eq:frame_dilaton}
\end{align}
We obtain the (higher-spin valued) effective metric 
\begin{align}
G^{\mu\nu} 
&= \frac{R}{{y^0}\alpha \beta^2 (y^0\beta)'}
\left( 
\begin{bmatrix}
 -\alpha^2 & \boldsymbol{0}^t \vspace{12pt} \\
\boldsymbol{0} & \beta^2 \one 
\end{bmatrix}
\right. \nn \\ &\qquad+  \left. 
\begin{bmatrix}
 0 &   \alpha \alpha' (\by\bu) \bu^t \vspace{12pt} \\
   \alpha \alpha' (\by\bu) \bu & 
 \big( y^0 \beta' \big(y^0 \beta'+ 2\beta \big) - ((\by\bu)\alpha')^2\big) \bu \bu^t
\end{bmatrix} 
\right).
\end{align} 
The frame $E^{\alpha\mu}$ has a spin-0 and a spin-2 part, which can easily 
be computed using \eqref{eq:S2_avg_tt}. From the latter equation, we obtain 
the pure spin-0 part of the frame,  
\begin{align}
[(E^{\alpha\mu})]_0 
&= \frac{y^0}{R} \begin{bmatrix}
-\alpha & \frac{ \alpha'}{3}  \by^t
\vspace{12pt}\\
\boldsymbol{0} & \big(\beta+\tfrac{{y^0}\beta'}{3}\big)   \one 
\end{bmatrix},\label{eq:avg_E}
\end{align}
while the pure spin-2 part is given by 
\begin{align}
[(E^{\alpha\mu})]_2 := (E^{\alpha\mu}) - [(E^{\alpha\mu})]_0 
= 
\frac{{y^0}}{R} \begin{bmatrix}
0 & \alpha' \big( (\by\bu) \bu^t -\tfrac{ \by^t}{3} \big)
\vspace{12pt}\\
\boldsymbol{0} & y^0\beta'  \big( \bu\bu^t - \frac{1}{3}\one\big) 
\end{bmatrix}.
\label{eq:E_spin2}
\end{align}
The effective metric has a spin-0, a spin-2 and a spin-4 part, which we will 
not work out explicitly here.

\subsection{Local normal coordinates}
\label{ssc:locnormcoord}

Due to the $\hs$ components, the physical meaning of a metric or frame is not clear 
{\em a priori}. As shown in \cite{Steinacker:2024unq}, it is always possible to 
locally eliminate these $\hs$ components, in local normal coordinates. We illustrate 
this in the following for the present backgrounds.

We fix the point $\xi=(\tau,0,0,0)$ as a 
reference point. We want to find local normal coordinates $\tilde{y}^\mu$ 
such that the frame has no higher-spin components at $\xi$ and satisfies
\begin{align}
\tilde{E}^{\mu\nu}|_{\xi} &= \{\tilde{t}^\mu,\tilde{y}^\nu\}|_\xi
=    \frac{\tau}{R} \begin{bmatrix} 
-\alpha(\tau) & \boldsymbol{0}^t \\ \boldsymbol{0} & \beta(\tau) \one
\end{bmatrix}.
\label{eq:frame_at_xi}
\end{align}

We make the ansatz 
\begin{align}
\tilde{y}^0 = y^0, \quad \tilde{y}^i = y^i + b(y^0) (\by\bu) u^i
= y^i + y^0 b(y^0)  u^0 u^i,
\end{align} 
which respects the local $SO(3)$. Note that we have $\by|_{\xi}=u^0|_{\xi}=0$.
This ansatz yields 
\begin{align}
\tilde{E}^{00} &= -\frac{y^0\alpha }{R},
\nn \\
\tilde{E}^{0j} 
&= \frac{1}{R} W\big( y^0(1+b), y^0\alpha  \big) \tfrac{(\by\bu)}{y^0}  u^j ,
\nn \\
\tilde{E}^{i0} &=0,
\nn \\
\tilde{E}^{ij} 
&= \frac{y^0}{R} \Big( \beta \delta^{ij}  +  \big( {y^0} \beta'
+ b ({y^0} \beta)' \big)  u^i u^j \Big)
\end{align}
where  $W$ is the Wronskian.
Because $\by|_{\xi}=0$, the requirement \eqref{eq:frame_at_xi} is fulfilled for
    \begin{align}
     &b(\tau) =-\frac{y^0\beta'}{( y^0\beta)'}\Bigg|_{y^0=\tau}.
    \label{eq:b_solu}
    \end{align} 
The local normal coordinates with $b$ as in \eqref{eq:b_solu} naturally extend for 
all $\tau>0$, i.e. on a time-like worldline. Then the frame $\tilde{E}^{\mu\nu}$ 
is globally given by
\begin{align}
(\tilde{E}^{\mu\nu}) 
&= (\overline{E}^{\mu\nu}) 
+\frac{\cW}{R}\begin{bmatrix}
0 &   \tfrac{(\by\bu)}{y^0} \bu^t  
\vspace{6pt}\\
\boldsymbol{0} & 0_{3\times 3}
\end{bmatrix}, \nn \\
\word{where}
(\overline{E}^{\mu\nu}) &=\frac{y^0}{R} \begin{bmatrix} 
-\alpha  & \boldsymbol{0}^t \\ \boldsymbol{0} & \beta  \one
\end{bmatrix}
\end{align}
and where the Wronskian $\cW = \cW(y^0) := W\big( \frac{y^0\beta}{(y^0\beta)'} ,y^0\alpha\big)$ 
determines the size of the higher-spin contributions. 
The determinant of the frame in local normal coordinates is then given by 
\begin{align}
\det\big(\tilde{E}^{\mu\nu}\big) 
&= -\frac{y^0\alpha (y^0\beta)^3 }{R^4}.
\label{eq:frame_det}
\end{align}

The symplectic density $\tilde{\rho}_M$ in the new coordinates reads 
\begin{align} 
\rho_M \dd^4 y = \tilde{\rho}_M \dd^4\tilde{y} 
= \frac{R}{y^0 \Big| \det \frac{\del \tilde{y}^\mu}{\del y^\nu} \Big|} \dd^4\tilde{y},
\end{align}  
and the Jacobian is readily computed to be 
\begin{align}\label{eq:Jacobian}
 \det \frac{\del \tilde{y}^\mu}{\del y^\nu} 
 &= 1 + b 
 = \frac{\beta}{(y^0\beta)'}
\qquad\Rightarrow \qquad 
\tilde{\rho}_M  
= R \frac{(y^0\beta)'}{y^0\beta}.
\end{align} 
The dilaton, which can be computed from the symplectic density and 
the determinant \eqref{eq:frame_det}, is not affected by the change 
of coordinates: we have 
\begin{align}
\tilde{\rho}^2=\rho^2
\end{align} 
 with $\rho^2$ as in 
\eqref{eq:frame_dilaton}. 
Consequently, the effective metric is obtained as 
\begin{align}
\tilde{G}^{\mu\nu} &= \frac{1}{R^2\tilde{\rho}^{2}} 
\left(  \begin{bmatrix}
    - (y^0\alpha)^2 & \boldsymbol{0}^t \\ \boldsymbol{0} & (y^0\beta)^2 \one
  \end{bmatrix} 
 +  \alpha \cW (\by\bu)
  \begin{bmatrix}
   0 &  \bu^t \vspace{4pt}\\
    \bu
   & -  \cW \tfrac{(\by\bu)}{\alpha{y^0}^2} \bu \bu^t
  \end{bmatrix}
\right), \label{eq:metric_hs}
\end{align}
so that 
\begin{align} 
\tilde{G}_{\mu\nu} 
&= \overline{G}_{\mu\nu} 
+  \frac{  (y^0\beta)'  \cW (\by\bu)}{R y^0}  \begin{bmatrix}
  \tfrac{\cW (\by\bu)  }{{y^0}^2\alpha} & \bu^t
  \vspace{4pt}\\ \bu & 0_{3\times3}
\end{bmatrix}, 
\nn \\ \word{where}
\overline{G}_{\mu\nu}
&=  \frac{y^0\alpha (y^0\beta)' }{R}
\begin{bmatrix}
- \frac{\beta^2}{\alpha^2} & \boldsymbol{0}^t \\ \boldsymbol{0} & \one
\end{bmatrix} \nn \\
\word{and} 
 \det \overline{G}_{\mu\nu} &= - \frac{ {y^0}^4 \alpha^2 \beta^2 ((y^0\beta)')^4}{R^4}.
\end{align}

If the higher spin entries of $\tilde{G}_{\mu\nu}$ are small in a 
neighborhood of $L=\{(y^0,\boldsymbol{0})\setassume y^0 \gr 0\}$, 
we may use the metric $\overline{G}_{\mu\nu}$ in this neighborhood.
The line element corresponding to $\overline{G}_{\mu\nu}$ is
\begin{empheq}[box=\widefbox]{align} 
\dd s^2_{\overline{G}} &= 
\frac{y^0 \beta^2 (y^0\beta)'}{R \alpha } \Big( -  \dd{y^0}^2
                                +  \tfrac{\alpha^2 }{\beta^2} \dd\by^2\Big).
\label{eq:line_element_locnormcoord}
\end{empheq}

To recognize this as a cosmological Friedman-Lemaitre-Robertson-Walker (FLRW) metric, 
we define the cosmological time $t$ via 
\begin{align}
t := \int_{0}^{y^0} 
    \Big( \frac{\beta(z)}{\alpha(z)} \frac{z\beta(z) (z\beta(z))'}{R} \Big)^{\frac12} \dd z,
\quad \dd t &= \Big(  \frac{\beta}{\alpha} \frac{y^0\beta (y^0\beta)'}{R} 
\Big)^{\frac12} \dd y^0.
\label{eq:def_t}
\end{align}
Expressed in terms of $t$, the line element \eqref{eq:line_element_locnormcoord} 
becomes 
\begin{align}
\dd s^2_{\overline{G}} 
&=  -  \dd t^2 +  a^2(t) \dd\by^2, 
\end{align}
where
\begin{align} 
a(t) =  \sqrt{ \frac{y^0 \alpha (y^0 \beta )' }{R}}
\end{align}
is the cosmic scale factor with $y^0=y^0(t)$,
 where the prime indicates a derivative with respect to $y^0$ as usual. 
If we indicate the derivative with respect to $t$ by a dot, then the Hubble 
parameter $H$ and the scalar curvature $\cR$ are given by 
\begin{align}
H &= \frac{\dot{a}}{a} = \frac{\dot{y}^0}{2} 
    \frac{\big(  y^0 \alpha (y^0 \beta )' \big)'}{  y^0 \alpha (y^0 \beta )' }; \nn \\
\cR &= \frac{\ddot{a}}{a} + H^2 = \frac{\ddot{y}^0 \big(  y^0 \alpha (y^0 \beta )' \big)' 
    + {\dot{y^0}}^2 \big(  y^0 \alpha (y^0 \beta )' \big)'' }{2 y^0 \alpha (y^0 \beta )'},
\label{eq:Hubble_Ricci}
\end{align}
where we always assume $y^0=y^0(t)$. In Subsection \ref{ssc:scales}, we will 
determine the asymptotic behavior of $H$ and $\cR$ at late times assuming that 
$\alpha$ and $\beta$ grow polynomially.

Let us estimate the magnitude of the higher spin-components of the frame 
$\tilde{E}^{\mu\nu}$ and metric $\tilde{G}^{\mu\nu}$ in local normal 
coordinates away from the line $L$. The only entries with higher-spin 
components are $\tilde{E}^{0j}$, and we have 
\begin{align}
\big| \tilde{E}^{0j} \big| 
&\leq \Bigg| \frac{\cW}{\sqrt{R {y^0}^3\alpha (y^0\beta)' }}\Bigg| 
\norm{\by}_{\overline{G}}, 
\label{eq:frame_hs_est}
\end{align}
where 
\begin{align}
 \norm{\by}_{\overline{G}} \,&\!:= \big( \overline{G}_{ij} y^i y^j \big)^{\frac12}
=\Big| \frac{y^0\alpha (y^0\beta)'}{R}  \delta_{ij} y^i y^j \Big|^{\frac12} 
\end{align}
is the length of $\by$ measured in terms of the metric $\overline{G}$.
 Similar estimates on the higher spin components of the metric yield 
\begin{align}
\big| \tilde{G}_{00} - \overline{G}_{00} \big| 
&\leq  \frac{ \cW^2}{  {y^0}^4 \alpha^2 }  \norm{\by}_{\overline{G}}^2 ; \nn \\ 
\big| \tilde{G}_{0j} - \overline{G}_{0j} \big| 
&\leq \Big| \frac{(y^0\beta)' \cW^2 }{R {y^0}^3\alpha}  \Big|^{\frac12} 
\norm{\by}_{\overline{G}} .
\label{eq:metric_hs_est}
\end{align}

 The higher-spin components of the metric and frame are negligible in the 
 semi-classical regime if the right-hand sides of \eqref{eq:frame_hs_est} 
 and \eqref{eq:metric_hs_est} become small compared to the reference frame 
 $\overline{E}^{\mu\nu}$ and metric $\overline{G}_{\mu\nu}$. That is, 
 \begin{align}
 \Delta \tilde{E}^{0j} &:= \frac{\big| \tilde{E}^{0j} \big|
 }{\big| \det \overline{E}^{\mu\nu} \big|^{\frac{1}{4}}} 
\leq \Bigg| \frac{ R \cW^2 }{ {y^0}^{5} (\alpha \beta)^{\frac{3}{2}} (y^0\beta)'} 
\Bigg|^{\frac12} \norm{\by}_{\overline{G}}; \nn \\
\Delta \tilde{G}_{00} &:= \frac{ \big| \tilde{G}_{00} - \overline{G}_{00} \big|
}{\big|\det \overline{G}_{\mu\nu} \big|^{\frac{1}{4}}}
 \leq \Bigg|  \frac{R \cW^2}{{y^0}^{5} \alpha^{\frac{5}{2}}\beta^{\frac{1}{2}} 
 (y^0\beta)'}\Bigg| 
    \norm{\by}_{\overline{G}}^2   ;
\nn \\
\Delta \tilde{G}_{0j} &:= \frac{\big| \tilde{G}_{0j} - \overline{G}_{0j} \big|
}{\big|\det \overline{G}_{\mu\nu} \big|^{\frac{1}{4}}}
\leq \Bigg| \frac{ R\cW^2 }{{y^0}^{5}\alpha^2 \beta (y^0\beta)' }   
\Bigg|^{\frac12} \norm{\by}_{\overline{G}}  
\label{eq:metric_frame_hs_deviation}
\end{align}
 must become small in a local patch at late times. Therefore, the bounds 
 \eqref{eq:metric_frame_hs_deviation} can be used to estimate the size of 
 the patch in which the semi-classical approximation is valid.

\subsection{Approximate diffeomorphisms relating different backgrounds}
\label{ssc:approx_symplectomorph}
In this section, we show that the gauge transformation 
\eqref{eq:gauge_def} corresponds to an approximate diffeomorphism
on $\cM^{1,3}$, under mild assumptions.

For $\Psi\in\cC(\cM)$, we introduce the notation $\delta\Psi := \{\Lambda(y^0) t^0, \Psi\}$. 
If $\Phi$ is a scalar function of the variables $y^\mu$, then its variation 
$\delta\Phi = \xi^\mu \del_\mu \Phi$ is described by the higher-spin valued vector 
\begin{align}
\xi^\mu = \frac{y^0}{R}\big( -\Lambda , \Lambda' (\by\bu) \bu\big),
\end{align}
which contains a classical (spin-0) and a spin-2 part, $\xi^\mu=[\xi^\mu]_0+[\xi^\mu]_2$, 
\begin{align} 
[\xi^\mu]_0 &= \frac{y^0}{R}\big( -\Lambda , \tfrac{\Lambda'}{3} \by\big),
\nn \\ \label{eq:avg_xi}
[\xi^\mu]_2 &= \frac{{y^0} \Lambda'}{R}\big(0,    (\by\bu) \bu - \tfrac{\by}{3}\big) .
\end{align}

The classical part describes a translation along the time direction $y^0$ near $\by=0$. 
Note that a time translation cannot implement the gauge transformation globally 
because the vector field is volume-preserving in the sense  
\begin{align}
\del_\mu (\rho_M [\xi^\mu]_0) = 0 .
\end{align}
Now we want to check the Lie derivatives of the frame \eqref{eq:hs_E_deviation}.
Using the explicit form of the spin-2 part of the frame given by 
\eqref{eq:E_spin2}, we find 
\begin{align}
 \{\tilde{t}^\alpha, [\xi^0]_2\} - \delta [E^{\alpha0}]_2 &=0; \nn \\
 \{\tilde{t}^0, [\xi^j]_2\} - \delta [E^{0j}]_2 
 &=- \frac{{y^0} y_k}{R^2} \big( y^0 W(\alpha,\Lambda)\big)' [u^k u^j]_2 \quad \in \cC^2; \nn \\
 \{\tilde{t}^i, [\xi^j]_2\} - \delta [E^{ij}]_2 
 &= \frac{2 {y^0}^3\Lambda'\beta'}{3R^2} u^i u^j + \frac{{y^0}^2}{R^2} \big( W(\beta,\Lambda) 
 + 2 y^0 \Lambda \beta'' \big) [u^i u^j]_2.
\end{align}
Thus, only the purely spatial part contributes to the right-hand side of 
\eqref{eq:hs_E_deviation}, which parametrizes the deviation from a 
proper transformation behavior of the frame under diffeomorphisms:
\begin{align}
 \cL_{[\xi]_0} [E^{0\mu}]_0-\big[ (\delta E^0)[y^\mu] \big]_0  
 &= \cL_{[\xi]_0} [E^{i0}]_0-\big[ (\delta E^i)[y^0] \big]_0 
 = 0 ; \nn \\
\cL_{[\xi]_0} [E^{ij}]_0-\big[ (\delta E^i)[y^j] \big]_0 
&=  \frac{2{y^0}^3}{9R^2}  \beta'\Lambda'\delta^{ij}. 
\label{eq:deviation_no_LNC}
\end{align}
The relative error \eqref{eq:relative_deviation_symplecto} becomes
\begin{align}
\Bigg|\frac{\cL_{[\xi]_0} [E^{ij}]_0 -\big[  (\delta_\Lambda E^i)[y^j]  \big]_0 
}{\cL_{[\xi]_0} [E^{ij}]_0} \Bigg|
&=\frac{ \frac{2}{9}  \big|\frac{y^0\beta'}{(y^0\beta)'} \big| \; \big| \frac{\Lambda'}{\Lambda}\big|}{
  \Big| 1 + \frac{y^0 \Lambda'}{3\Lambda}  + \frac{y^0 (y^0\beta)''}{3 (y^0\beta)'}
  - \frac{2}{9} \frac{y^0\beta'}{(y^0\beta)'} \frac{\Lambda'}{\Lambda} \Big|}.
  \label{eq:beta_dev}
\end{align}

Suppose that $\beta\sim {y^0}^\mu$ as $y^0\gg0$ for some $\mu\in\bR\setminus\{-1\}$. Then 
\begin{align}
\Bigg|\frac{\cL_{[\xi]_0} [E^{ij}]_0 -\big[  (\delta_\Lambda E^i)[y^j]  \big]_0 
}{\cL_{[\xi]_0} [E^{ij}]_0} \Bigg|
&= \frac{2}{9}  
  \frac{ \big|\frac{\mu }{\mu+1} \big| \; \big| \frac{\Lambda'}{\Lambda}\big|}{
  \Big| 1 + \frac{y^0 \Lambda'}{3\Lambda}  + \frac{\mu}{3 }
  - \frac{2}{9} \frac{\mu}{\mu+1} \frac{\Lambda'}{\Lambda} \Big|}. 
  \label{eq:beta_s_dev}
\end{align}
Therefore, any $\Lambda\sim {y^0}^a$ as $y^0\gg0$ will make the error decay 
as $y^0\to\infty$ \footnote{With the exception $\Lambda=c {y^0}^{-3-\mu}$.}, 
and the frame transforms as expected.

Suppose we are given a fixed gauge configuration $(\alpha,\beta)$. In order to 
determine which other gauge configurations $(\tilde{\alpha},\tilde{\beta})$ lie 
in the same gauge orbit as $(\alpha,\beta)$, one needs to compute the finite gauge 
transformations implemented by $\Lambda(y^0)t^0$ such that \eqref{eq:beta_dev} 
becomes small at late times. This is beyond the scope of our paper, but our 
evaluation of physical consistency conditions in Subsection \ref{ssc:scales} below 
suggests that the ratios $\tfrac{\alpha}{\beta}$ and $\tfrac{\beta}{\alpha}$ must 
not grow too fast. That is to say, $\alpha$ and $\beta$ should not behave too differently. 
We will describe more concrete bounds on their behavior in the following two subsections.

\subsection{Physical scales and consistency conditions}
\label{ssc:scales}
Let us now explicitly work out constraints on the background $\tT^{\mu}$
so that the conditions for reasonable semi-classical physics described 
in Subsection \ref{ssc:scales_intro} are fulfilled. We restrict our 
considerations to backgrounds where $\alpha$ and $\beta$ (and their 
derivatives) behave monotonously like powers in the late time regime, 
\begin{align}
 \alpha\sim {y^0}^{\la}, \quad \beta\sim {y^0}^{\mu} \word{for} y^0\gg0.
\end{align}

On the non-commutative level, the leading contribution to $\abs{[Y^\mu,Y^\nu]
[Y_\mu,Y_\nu]}$ is the mixed spatial-temporal entry \cite{Steinacker:2024unq}.
This should remain true on the semi-classical level, which is easy to verify
if the ratio $\tfrac{\alpha}{\beta}$ does not grow too 
fast. In this case, we generically find 
\begin{align}
 L_{\rm NC}^{\overline{G}} 
 &\sim 2 \abs{ \overline{G}_{00} \overline{G}_{ij} \{y^0,y^i\} \{y^0, y^j\}}^{\frac{1}{4}}
           = \sqrt{ \frac{4r}{R} \beta  (y^0\beta)'}  \; y^0 \sim {y^0}^{1+\mu}.
\end{align}

The cosmological time $t$ defined in \eqref{eq:def_t} and the derivatives 
of $y^0$ with respect to $t$ satisfy
\begin{align}
 t 
 &\sim \frac{2}{3\mu-\la+3} {y^0}^{\frac{3\mu-\la+3}{2}}, 
 \quad 
 &y^0\sim \Big(\frac{3\mu-\la+3}{2} t\Big)^{\frac{2}{3\mu-\la+3}},
 \quad \nn \\
 \dot{y}^0 
 &\sim {y^0}^{\frac{-3\mu+\la-1}{2}},
 &\ddot{y}^0 
  \sim  \frac{-3\mu+\la-1}{2}  {y^0}^{-3\mu+\la-2}.
\end{align} 
Consequently, the scale factor, Hubble rate an Ricci curvature behave like 
\begin{align}
 a &\sim {y^0}^{\frac{1+\la+\mu}{2}} \sim t^{\frac{1+\la+\mu}{3\mu-\la+3}}; \nn \\
 H &\sim  \frac{1+\la+\mu}{2} {y^0}^{\frac{-3\mu+\la-3}{2}} 
 \sim \frac{1+\la+\mu}{3\mu-\la+3}  \;\frac{1}{t}; \nn \\
 \cR &\sim  - \frac{ (1+\la+\mu)(\mu-3\la+1)   
    }{4  } {y^0}^{-3\mu+\la-3} \nn \\
    &\sim   - \frac{ (1+\la+\mu)(\mu-3\la+1)   
    }{ (3\mu-\la+3)^2 } \;\frac{1}{t^2}. 
    \label{eq:HR}
\end{align}

In a physically reasonable situation, the minimum requirements should be that 
the Hubble rate and Ricci curvature  decay with $y^0$, and that the cosmic scale 
$L_{\rm cosm}$ is large compared to the non-commutativity scale 
$L_{\rm NC}^{\overline{G}}$, 
\begin{align}
 \frac{L_{\rm cosm}}{L_{\rm NC}^{\overline{G}}} 
 = \frac{1}{H L_{\rm NC}^{\overline{G}}}\sim 
 {y^0}^{\frac{1+\mu-\la}{2}} \gg 1 
 \word{for} {y^0}\gg0.
 \label{eq:Lcosm_vs_LNC}
\end{align}
The decay of the Hubble rate implies $\la\sm 3(1+\mu)$, which
is equivalent to $t$ increasing with $y^0$. The requirement that the cosmic 
scale becomes large compared to the scale of non-commutativity implies $\la\sm 1+\mu$. 
Both constraints are upper bounds on the growth of the ratio $\tfrac{\alpha}{\beta}$, 
as expected.

The second bound $\la\sm1+\mu$ also guarantees that the scale of non-commutativity is 
dominated by the mixed spatial-temporal entries, as should be:
\begin{align}
       \left| \frac{G_{ij}G_{kl}\theta^{ik}\theta^{jl}}{G_{ij}G_{00}\theta^{i0}\theta^{j0}}
       \right|
       \sim {y^0}^{-2(1+\mu-\la)}.
\end{align}

Finally, the estimates \eqref{eq:metric_frame_hs_deviation} for the 
higher-spin components of the frame and metric read 
\begin{align}
 \Delta \tilde{E}^{0j} &:= \frac{\big| \tilde{E}^{0j} \big|
 }{\big| \det \overline{E}^{\mu\nu} \big|^{\frac{1}{4}}} 
\leq \Bigg| \frac{\sqrt{R} \lambda}{(1+\mu)^{\frac32}} \Bigg| {y^0}^{-\frac{6+5\mu-\la}{4}}
\norm{\by}_{\overline{G}}; \nn \\
\Delta \tilde{G}_{00} &:= \frac{ \big| \tilde{G}_{00} - \overline{G}_{00} \big|
}{\big|\det \overline{G}_{\mu\nu} \big|^{\frac{1}{4}}}
 \leq \Bigg|  \frac{R \la^2}{(1+\mu)^3}\Bigg| 
    {y^0}^{-\frac{6+3\mu+\la}{2}} \norm{\by}_{\overline{G}}^2    ;
\nn \\
\Delta \tilde{G}_{0j} &:= \frac{\big| \tilde{G}_{0j} - \overline{G}_{0j} \big|
}{\big|\det \overline{G}_{\mu\nu} \big|^{\frac{1}{4}}}
\leq \Bigg| \frac{ \sqrt{R}\la }{(1+\mu)^{\frac32} }   
\Bigg| {y^0}^{-\frac{3+2\mu}{2}}\norm{\by}_{\overline{G}}  . \label{eq:hs_estimates}
\end{align}

These bounds can be used 
to estimate the size of the patch at late times in which the semi-classical 
approximation is valid: if all powers of $y^0$ on the right-hand side are 
negative, this patch will be large at late times. However, if at least one 
estimate grows with $y^0$, then the semi-classical approximation is only 
valid in a small patch. 

Finally, we require the dilaton to be either constant or increasing with $y^0$, 
corresponding to weak Yang-Mills coupling $g_{\rm YM} \sim \rho^{-1}$ \cite{Steinacker:2024huv}.
Using the formula \eqref{eq:frame_dilaton}, this means that $3(1+\mu)+\la\geq0$. 
This condition might not be strictly necessary, but we will 
find later that it only excludes some \emph{shrinking} 
FLRW spacetimes, which are anyway not interesting from a physical point of view.

Let us summarize all physical requirements: 
\begin{empheq}[box=\widefbox]{align}
 &\text{$t$ grows, $H$ and $\cR$ decay with $y^0$:}  &&\la\sm 3(1+\mu), \nn \\
 &L_{\rm cosm} \gg L_{\rm NC}: &&\la\sm 1+\mu, \nn \\
 &\text{semi-class.~approx. valid in large patch:} && \begin{cases}
 \la\sm 6+5\mu,  \\
 \la\gr -6-3\mu, \\ 
 \mu\gr-\frac{3}{2},
 \end{cases} \nn \\
 &\text{dilaton constant or growing with $y^0$:} && \la\geq-3(1+\mu).
 \label{eq:physical_growth_cond}
\end{empheq} 
These conditions can easily be satisfied simultaneously. For example, 
the undeformed background $\lambda = \mu = 0$ and small deformations 
thereof satisfy \eqref{eq:physical_growth_cond}.

\subsection{Distinguished gauges}
\label{ssc:canon_gauge} 
Let us now consider special choices of one of the two parameters. We investigate 
the \emph{timelike gauge} $\beta\equiv 1$ and the \emph{covariant gauge}  
$\alpha\equiv\beta$. 
We will verify explicitly that these different gauges lead to the same classical 
geometry for some specific backgrounds, by computing the effective metric in local 
normal coordinates. This validates in particular the physical significance of local 
normal coordinates.

\subsubsection{Timelike gauge}
As outlined in Subsection \ref{ssc:locnormcoord}, things become particularly 
simple in the gauge $\beta\equiv1$ because then, the local normal coordinates 
coincide with the original coordinates $y^\mu$. In this gauge, we have 
\begin{align}
 \Box_{\tT} \tT^0 =0 \word{and}
 \Box_{\tT} \tT^j = \frac{\alpha}{R^2 } \big(  Y^0 \alpha \big)'   \tT^j .
\label{eq:BoxU_U_beta1}
\end{align}

On the semi-classical level, the frame and its determinant are given by  
\begin{align}
(E^{\alpha\mu}) 
&= \frac{y^0}{R} \begin{bmatrix}
-\alpha & \alpha' (\by\bu) \bu^t \\
\boldsymbol{0} &  \one
\end{bmatrix}, \quad 
\det(E^{\alpha\mu}) 
= -\frac{{y^0}^4\alpha }{R^4},
\end{align} 
and the effective metric reads 
\begin{align}
G_{\mu\nu} 
&= \overline{G}_{\mu\nu}
+  \frac{{y^0}\alpha'}{R} (\by\bu)
\begin{bmatrix}
 \frac{ \alpha' (\by\bu)}{ \alpha} & \bu^t  \\
 \bu & 0_{3\times3}
\end{bmatrix} 
,\quad 
 \overline{G}_{\mu\nu} =  \frac{y^0\alpha}{R}
 \begin{bmatrix}
- \frac{1}{\alpha^2} & \boldsymbol{0}^t  \\
\boldsymbol{0} &   \one
\end{bmatrix}.
\end{align}

If we assume a powerlike behavior $\alpha=const.\times {y^0}^\la$, the 
physical consistency conditions \eqref{eq:physical_growth_cond} in the 
timelike gauge ($\mu=0$) become 
\begin{align}
-3\leq\la\sm1.
\end{align}
Defining $t := const.\times {y^0}^{\frac{3-\la}{2}}$, the line 
element associated to $\overline{G}_{\mu\nu}$ becomes 
\begin{align}
\dd s_{\overline{G}}^2 
&= const. \times \Big( - \dd{t}^2 + {t}^{2 \frac{1+\la}{3-\la}}\dd\by^2\Big),
\label{eq:line_element_simple1}
\end{align}
which is the line element of a spatially flat FLRW spacetime with 
scale factor $a_\la(t)=t^{\frac{1+\la}{3-\la}}$. Different values of 
$\la$ yield differently behaving scale factors: 
\begin{align}
\la\in (-1,1): \quad &\textup{expanding FLRW}; \nn \\
\la=-1:\quad &\textup{static case}; \nn \\
\la\in[-3, -1):  \quad &\textup{shrinking FLRW}. \nn 
\end{align}
The scale factor approaches linear growth as $\la\nearrow1$, and the 
dilaton \eqref{eq:frame_dilaton} satisfies 
\begin{align}
\tilde{\rho}^2 \sim {y^0}^{3+\la} \sim t^{\frac{6+2\la}{3-\la}}.
\end{align}

Let us mention two distinguished backgrounds.

\paragraph{(i) The classical solution.} 
By \emph{classical solution}, we mean the solution that makes $\tT^\mu$ a 
solution of the equation $\Box_{\tT} \tT^\mu=0$. In the current gauge, this
means that $\alpha=\tfrac{const.}{y^0}$. If we define $t:=const.\times {y^0}^2$, 
the line element \eqref{eq:line_element_simple1} becomes the line element of the 
Minkowski metric,
\begin{align}
\dd s_{\overline{G}}^2 
&= const. \times \Big( - \dd t^2 + \dd\by^2\Big).
\end{align}
However, the dilaton is not constant, but grows like $\tilde{\rho}^2 \sim   t$.

\paragraph{(ii) The constant dilaton background.} 
 The dilaton is constant for $\alpha=\tfrac{const.}{{y^0}^3}$. Defining 
 $t:= const.\times {y^0}^3$, the line element 
 \eqref{eq:line_element_simple1} becomes
\begin{align}
\dd s_{\overline{G}}^2 
&=const. \times \Big( - \dd{t}^2 + {t}^{-\frac{2}{3}} \dd\by^2\Big).
\end{align}

\subsubsection{Covariant gauge}
Let us now discuss the covariant gauge $\alpha\equiv\beta$. 
In this gauge, the frame is given by 
\begin{align}
(\tilde{E}^{\mu\nu}) 
&= (\overline{E}^{\mu\nu}) 
+  \frac{(y^0\beta)^2 (y^0\beta)''}{R ((y^0\beta)')^2}\begin{bmatrix}
0 &   \tfrac{(\by\bu)}{y^0} \bu^t  \\
\boldsymbol{0} & 0_{3\times 3}
\end{bmatrix}, \quad 
\overline{E}^{\mu\nu} =\frac{y^0\beta}{R} \eta^{\mu\nu},
\end{align} 
and the effective metric reads  
\begin{align}
\tilde{G}_{\mu\nu} 
&= \overline{G}_{\mu\nu} 
+  \frac{ y^0\beta^2 (y^0\beta)''  (\by\bu)}{ R (y^0\beta)'}  \begin{bmatrix}
    \frac{ \beta (y^0\beta)''  (\by\bu)}{y^0((y^0\beta)')^2}  
     & \bu^t \\ \bu & 0_{3\times3}
\end{bmatrix},  \nn \\
 \word{where}
\overline{G}_{\mu\nu} &=  \frac{ ((y^0\beta)^2)'}{2 R} \eta_{\mu\nu}.
\end{align} 

If we assume a powerlike behavior $\beta=const.\times {y^0}^\mu$, the  
physical consistency conditions \eqref{eq:physical_growth_cond} in the 
timelike gauge ($\la=\mu$) reduce to the constraint
\begin{align}
 \mu\geq-\frac{3}{4}.
\end{align}
 Then we define $t:= const.\times {y^0}^{\mu+\frac32}$, so that the line element becomes 
\begin{align}
\dd s_{\overline{G}}^2 
= const.\times  \big(-\dd t^2 + t^{\frac{4\mu+2}{2\mu+3}} \dd\by^2 \big).
\end{align}
The scale factor of this FLRW line element is $a_{\mu}(t)= t^{\frac{2\mu+1}{2\mu+3}}$, 
and in complete analogy to the timelike gauge, different values of $\mu$ 
yield differently behaving scale factors: 
\begin{align}
\mu\in (-\tfrac{1}{2},\infty): \quad &\textup{expanding FLRW}; \nn \\
\mu=-\tfrac{1}{2}:\quad &\textup{static case}; \nn \\
\mu\in[-\tfrac{3}{4}, -\tfrac{1}{2}):  \quad &\textup{shrinking FLRW}. \nn 
\end{align}
The scale factor approaches linear growth as $\mu\to\infty$, 
and the dilaton \eqref{eq:frame_dilaton} satisfies
\begin{align}
\tilde{\rho}^2 \sim {y^0}^{3+4\mu} \sim t^{\frac{6+8\mu}{3+2\mu}}.
\end{align}

Let us again discuss some distinguished solutions. 

\paragraph{(i) The classical solution.}
For the dynamical background to satisfy the equations $\Box_{\tT} \tT^\mu=0$, 
$\beta$ must be of the form 
\begin{align}
 \beta(y^0) = \frac{\sqrt{cy^0+d}}{y^0}
\end{align}
for some constants $c,d$. We assume $c\gr 0$ to exclude the degenerate 
solution and $d\geq0$. For large times, we have $\beta\sim {y^0}^{-\frac12}$. 

As in the timelike gauge, the metric $\overline{G}$ is simply a multiple of 
the Minkowski metric -- now even without a change of time variable, 
\begin{align}
\dd s^2_{\overline{G}} 
&= const.\times
\Big( -  \dd{y^0}^2 +   \dd\by^2\Big).
\end{align}
As in the timelike gauge, the dilaton grows like $\tilde{\rho}^2 = t$.

\paragraph{(ii) The constant dilaton background.}
The dilaton is constant for $\beta$ of the form 
\begin{align}
 \beta(y^0)=\frac{(cy^0+d)^{\frac{1}{4}}}{y^0}
\end{align}
If we define $t:= const.\times (cy^0+d)^{\frac{3}{4}}$, then, 
again as in the timelike gauge, the line element corresponding 
to such $\beta$ is given by 
\begin{align}
\dd s^2_{\overline{G}} 
&= const.\times \Big( -\dd{t}^2 +t^{-\frac{2}{3}} \dd\by^2\Big),
\end{align}
which describes the same shrinking, spatially flat FLRW spacetime as 
the constant dilaton solution in the timelike gauge.

\paragraph{(iii) Background with exactly vanishing $\hs$ components.}
Consider $\beta= \frac{d}{y^0}+c$ for some $c\gr 0$. The corresponding background 
is a sum of the reference background $T^\mu$ and the background $U^\mu$ (which on 
its own yields a degenerate frame and metric). 
For this background, the higher-spin 
components of the frame $\tilde{E}^{\mu\nu}$ and metric $\tilde{G}^{\mu\nu}$
 in local normal coordinates 
vanish exactly because $(y^0\beta)''=0$. Moreover, the frame and metric are 
non-degenerate because $(y^0\beta)'=c\neq0$. 

Thus, the transition to local normal coordinates removes the higher-spin components 
globally for such $\beta$, and not only along the line $L$. We have
\begin{align}
 \tilde{G}_{\mu\nu} = \overline{G}_{\mu\nu},
\end{align}
and the FLRW line element is the same as for the reference background, 
where $a(t)=t^{\frac{1}{3}}$ with $t=const.\times (cy^0+d)^{\frac{3}{2}}$.

\begin{remark}
 Note that by \eqref{eq:metric_hs}, a background where the $\hs$ 
 components of the metric vanish exactly can be found for any gauge 
 configuration $\alpha=f(\beta)$ such that 
 \begin{align}
  \cW = W\Big( \frac{y^0\beta}{(y^0\beta)'}, y^0 f(\beta) \Big) =0. 
  \label{eq:W_hs_vanish}
 \end{align}
  We presented this background in more detail  for the covariant gauge 
  because then \eqref{eq:W_hs_vanish} becomes particularly simple: 
  $(y^0\beta)''=0$, $(y^0\beta)'\neq0$.
\end{remark}

\section{FLRW quantum spacetimes and the IKKT model}
\label{sec:IKKT}

Let us finally describe how the cosmological quantum spacetimes 
described above could be incorporated into the IKKT model\footnote{It is sometimes 
questioned whether the IKKT mode admits any non-trivial backgrounds or saddle points 
with reduced dimension. These doubts are resolved by the example of flat $D3$ BPS 
branes realized by the Moyal-Weyl quantum plane $\R^{3,1}_\theta$, leading to 
noncommutative $\cN=4$ SYM \cite{Aoki:1999vr}.}. To this end, 
consider the bosonic part of the IKKT action \eqref{eq:IKKT_S}, 
\begin{align}
S[T] := \Tr\big( [T^a,T^b] [T_a,T_b] \big), \quad a,b=0,\dots,9.
\end{align}
We split the $T^a$ into the matrices parametrizing the background and the 
extra dimensions, respectively,
\begin{align}
T^\mu=\tT^\mu, \quad T^I= \Phi^I,\quad \mu=0,\dots,3,\quad I=4,\dots,9.
\end{align}
Then 
\begin{align}
S[T] 
&= -\Tr\Big( \tT^\nu \big( \Box_{\tT}  + \Box_\Phi  \big)\tT_\nu
+ \Phi^J \big( \Box_{\tT} + \Box_\Phi \big) \Phi_J \Big),
\end{align}
and the equations of motion for the background and extra dimensions become 
\begin{align}
 \big( \Box_{\tT}  + \Box_\Phi  \big) T^a &=0, \quad a=0,\dots,9. 
\label{eq:eom_10d}
\end{align}
Taking into account \eqref{eq:BoxU_U}, the background $\tT^\mu$ and the 
non-trivial extra-dimensions $\Phi^I$ satisfy the \emph{classical} IKKT equations 
of motion if and only if
\begin{align}
\Box_\Phi \tT^0 &=0; \nn \\
\Box_\Phi \tT^j &= -\frac{\alpha}{R^2\beta} \big(Y^0\alpha (Y^0\beta)'\big)' \tT^j; \nn\\
 (\Box_{\tT}+ \Box_\Phi ) \Phi^I &= 0 . \label{eq:IKKT_BoxPhi_U}
\end{align}
Since $\Box_\Phi=[\Phi^I, [\Phi_I,\cdot]]$ is a positive operator, non-trivial 
solutions for the $\Phi^I$ are possible only if $\Box_{\tT} \Phi^I\neq0$.
Following \cite{MantaStein_deformed}, the latter is naturally achieved by 
introducing some time-dependence to the internal generators $\Phi_I$, corresponding 
to a non-trivial $R$ charge. This is achieved by an ansatz of the form
\begin{align}
T_I^+ &:= \chi(Y^0) \ee^{\ii\omega(Y^0)} (\cK_{2I} + \ii \cK_{2I+1}), \nn\\
T_I^- &:=\chi(Y^0) \ee^{-\ii\omega(Y^0)} (\cK_{2I} - \ii \cK_{2I+1}), 
\quad I=2,3,4, 
\label{eq:ansatz_extradim}
\end{align}
where the $\cK_J$ are fixed matrices. This respects the full $E(3)$ symmetry of 
the cosmology, and it allows to stabilize a large hierarchy between the cosmic 
IR scale and a UV scale set by the Kaluza-Klein modes arising on the compact 
space $\cK$ generated by the $\Phi^j$, as demonstrated for $k=-1$ in 
\cite{MantaStein_deformed}. However, this will not affect the equations of 
motion for the background $\tT^\mu$.

To find physically reasonable solutions for $\tilde T^j$, it thus appears that 
quantum corrections are required. Indeed, quantum corrections arising e.g. from 
the one-loop effective action will certainly modify the equations of motion, and
a partial analysis of such effects was carried out for the $k=-1$ case in 
\cite{Battista:2023glw,MantaStein_deformed}. Using covariance arguments, one may 
reason along the lines of \cite{MantaStein_deformed} that the one-loop quantum 
corrections modify the classical equations of \eqref{eq:IKKT_BoxPhi_U}, so that the right-hand 
side is not vanishing but dependent on the dilaton and the scale $m_{\cK}^2$ of the extra 
dimension, yielding in particular
\begin{align}
 -\frac{\alpha}{R^2\beta} \big(Y^0\alpha (Y^0\beta)'\big)' \sim F(\tilde{\rho}^2) m_{\cK}^2 
 + \text{higher-loop corrections},
\end{align}
where $F(\tilde{\rho})^2$ is a function of the dilaton. However, a detailed study of this 
issue  is beyond the scope of this paper.

Let us also mention that the one-loop effective action on covariant spacetimes arising 
from the IKKT model was computed in \cite{Steinacker:2023myp} and worked out in some detail 
for cosmological $k=-1$ quantum spacetimes in \cite{Battista:2023glw} under the assumption 
that the full ten-dimensional background has a product structure $\cM^{1,3}\times\cK$. 
For related works, see \cite{Manta:2024vol, Steinacker:2024unq, Battista:2022hqn, 
Battista:2022vvl}.

\section{Discussion}

We have described a large class of spatially flat cosmological quantum spacetimes 
arising from the doubleton representations of $\mso(4,2)$. The obtained spacetimes 
can be expanding with a Big Bang geometry or shrinking with a Big Crunch geometry. 
We have elaborated on the algebraic and semi-classical structure of these spacetimes 
and the associated matrix configurations, including the appearing higher-spin 
degrees of freedom, the effective metric and geometry, the definition of appropriate 
derivation and the form of local normal coordinates which make the higher-spin components 
of the effective metric vanish along a timelike curve. 

A particular focus was on the relation between gauge transformations of the matrix model 
and approximate diffeomorphisms on the spacetime. We elaborate two distinguished gauges, the 
timelike gauge and the covariant gauge. Among other results, we have shown 
that particular backgrounds -- the ones solving the classical equations of 
motion and the ones yielding a constant dilaton -- yield equivalent geometries in these 
two gauges. Determining the finite gauge transformations relating arbitrary gauges is 
however beyond the scope of this work. 

An interesting result is the existence of a background $\beta=c+\tfrac{d}{y^0}$ in the 
covariant gauge, where  the higher-spin contributions to the effective metric 
can be eliminated globally 
by going to local normal coordinates. This background, which corresponds to  
an expanding FLRW spacetime with scale factor $a(t)\sim t^{\frac{1}{3}}$,
is a sum of the reference background investigated in \cite{Gass:2025tae} and the background 
$\beta_0\sim\tfrac{1}{y^0}$, which on its own yields a degenerate frame and (inverse) metric.

An important deviation from the case of cosmological $k=-1$ spacetimes is the observation 
that the background corresponding to a constant dilaton yields a shrinking spacetime, rather 
than an expanding spacetime. One might argue that the case of a 
constant dilaton is, in some sense, the "most physical". However, such a conclusion is premature 
without a more complete understanding of the resulting physics. Indeed, we also obtain 
a large class of expanding spacetimes with a Big Bang provided that the dilaton has some mild 
growing behavior during the cosmic evolution. This should ensure a weak coupling (at late times) 
for the low-energy gauge theory governing the internal fluctuation modes corresponding to $\cK$.

Finally in Section \ref{sec:IKKT}, we have outlined how the cosmological $k=0$ quantum spacetimes 
under consideration can be embedded into the full ten-dimensional IKKT model. The 
details of such an embedding are beyond the scope of the present paper. Although we expect 
that this works similar to the better-known case of $k=-1$ spacetimes \cite{Battista:2022vvl,Battista:2023glw,Battista:2022hqn,MantaStein_deformed}, 
there are some important differences. In particular, the fact that the $k=0$ case is 
"more commutative" means that time-dependent extra-dimensions will not affect the 
classical equations of motion of spacetime, so that quantum effects are required to stabilize spacetime.

\appendix

\section{Remarks on the quantization and the semi-classical correspondence}
\label{app:k-0}
We discuss the explicit quantization and the semi-classical correspondence for the symplectic space $\C P^{1,2}$ underlying the covariant quantum spacetime $\cM^{1,3}$.

Since $\C P^{1,2}$ is a coadjoint orbit of $SU(2,2)$, it carries a natural (Kirillov-Kostant-Souriau) invariant symplectic form. One then expects that a $SU(2,2)$-equivariant quantization in terms of operators on suitable representations exists. The appropriate representations are easily recognized as the doubleton representations, which can be constructed conveniently using an oscillator construction; this is discussed in detail in \cite{Manta:2025inq,Steinacker:2024unq}.
This provides a quantization map where  polynomial functions on $\C P^{1,2}$
generated by $m^{ab} \in \msu(2,2)^*$ are mapped to (symmetrized) polynomials in the generators $M^{ab}$ of $\msu(2,2)$ in the chosen representation, respecting the Poincare-Hilbert series i.e. the number of independent polynomials of any given degree. Using the oscillator construction and its Poisson version, it is then straightforward to verify \eqref{eq:Q_properties}.

Another -- often more useful -- way to define quantization is based on (quasi-)coherent i.e. optimally localized states $|x\rangle$ via
\begin{align}
\label{quant-map-coherent}
\Phi = \int \Omega \phi(x)|x\rangle\langle x|,
\qquad \phi(x) \approx \langle x|\Phi|x\rangle \ .
\end{align}
For compact coadjoint orbits of semi-simple Lie groups, such coherent states are obtained as group orbits of the highest weight state of some representation \cite{Perelomov:1986uhd}. This orbit is then isomorphic to the corresponding coadjoint orbit. For non-compact groups, the story is more complicated, and there is no obvious choice of coherent states. In the present case there appears to be a natural candidate given by the lowest weight state of the doubleton representations of $SU(2,2)$; however their group orbit is not isomorphic to $\C P^{1,2}$ but higher-dimensional, and moreover these states are not optimally localized. Nevertheless, it is possible to construct quasi-coherent states which are optimally localized in some local patch of $\C P^{1,2}$, i.e. saturating the uncertainty implied by the symplectic volume form $\Omega$;
we refer to \cite{Manta:2025inq} for details.
This allows to use \eqref{quant-map-coherent} in the local patch, which justifies the semi-classical treatment of $\cM^{1,3}$ in the present paper.

\section{Identities from \texorpdfstring{\boldmath{$k=-1$}}{} cosmological quantum spacetimes}
\label{app:k-1}
It was shown previously that a reference metric of a cosmological spacetime with $k=-1$ 
and a Big Bounce is induced by the static choice 
\begin{align}
 T^\mu_{k=-1} := \frac{1}{R} M^{\mu4},
\end{align}
which can be generalized and made dynamical by replacing $T^\mu_{k=-1}$
by $\tT^\mu_{k=-1} := f(X^4)T^\mu_{k=-1}$. Here, $f$ is a (well-behaved) function 
of $X^4$, which can be expressed in terms of the cosmological time variable only 
\cite{Sperling:2019xar,Sperling:2018xrm,MantaStein_deformed,Steinacker:2024unq}.
The $X^\mu$ and $T^\mu_{k=-1}$ satisfy the commutation relations
\begin{align}
[X^\mu,X^\nu] &= -\ii r^2 M^{\mu\nu}; \nn \\
  [T^\mu_{k=-1},T^\nu_{k=-1}] &= \frac{\ii}{R^2} M^{\mu\nu}; \nn \\
[T^\mu_{k=-1},X^\nu] &= \frac{\ii X^4}{R} \eta^{\mu\nu}. \label{eq:TX_commu}
\end{align}

The generators $M^{ab}$ and the matrices $X^a$ satisfy 
a number of $SO(4,1)$-invariant constraints in the doubleton 
representations. Among these, we will make use of the identity 
\eqref{eq:defX} and \cite{Steinacker:2024unq}
\begin{align}
  \eps_{abcde} M^{ab} M^{cd} = \frac{4n}{r} X_e,\quad a,b,c,d,e=0,...,4.
  \label{eq:epsMM}
\end{align}

For large $n\to\infty$, we have in the semi-classical limit $R \sim \tfrac{nr}{2}$, 
$x^\mu \sim X^\mu$, $t_{k=-1}^\mu\sim T_{k=-1}^\mu$, $m^{ab} \sim M^{ab}$ and 
$\{\cdot,\cdot\} \sim -\ii [\cdot,\cdot]$. Moreover, the Poisson-brackets with 
$x^4\sim X^4$ relate $x^\mu$ and $t_{k=-1}^\mu$, 
\begin{align}
\{x^4,x^\mu\} = r^2R t_{k=-1}^\mu \word{and} \{x^4,t_{k=-1}^\mu\} = \frac{ x^\mu}{R}.
\label{eq:Poisson_x4}
\end{align}

 A standard parametrization of the $x^a$ is 
 \begin{align}
 x^a &= R\begin{pmatrix}
  \cosh(\eta) \cosh(\chi) \\
  \cosh(\eta) \sinh(\chi) \boldsymbol{\Omega} \\
 \sinh(\eta)
 \end{pmatrix},
 \end{align}
where $\boldsymbol{\Omega}\in S^2\hookrightarrow \bR^3$, and where $\eta\in\bR$ 
parametrizes time and $\chi\in\bR$ is a spatial variable.

\section{Semi-classical generators of rotations and boosts}
\label{app:mmnunu}
We determine the semi-classical form of the generators of rotations, 
$M^{ij}\sim m^{ij}$ and of the boosts, $M^{i0}\sim m^{i0}$ as functions 
of the vectors $\by$ and $\bt$ and the $SO(3)$-invariant variables $y^0$, 
$t^0$ and $y:=\sqrt{\delta_{ij}y^iy^j}$, assuming that the parameter $n$ parametrizing 
the representation is either large, $n\gg0$, or vanishes $n=0$. 
Starting point are the concrete forms $m^{\mu\nu}_{k=-1}$ of the semi-classical 
generators of $SO(3,1)$ in the $k=-1$ case, which is in the large-$n$-limit given 
by \cite[Eq.~(5.4.13)]{Steinacker:2024unq}
\begin{align}
m^{\mu\nu} = - \frac{R}{R^2+{x^4}^2} \Big( 
x^4 (x^\mu t_{k=-1}^\nu - x^\nu t_{k=-1}^\mu ) 
+ R\eps^{\mu\nu\alpha\beta} x_\alpha t_{k=-1,\beta}
\Big), \quad n\gg0, \label{eq:mnunu-1}
\end{align} 
where $x^\mu=y^\mu - \delta^{\mu0} x^4$, $t_{k=-1}^\mu =t^\mu 
- \frac{1}{R} m^{\mu 0}$ is the standard background for a cosmological 
$k=-1$ quantum spacetime, related to $y^\mu$ via \eqref{eq:xy_rel}.

In the case $n=0$, we have the simpler form \cite{Manta:2025inq}
\begin{align}
m^{\mu\nu} &= \frac{R}{x^4} ( t_{k=-1}^\mu  x^\nu- t_{k=-1}^\nu x^\mu). 
\label{eq:mmunu_minimal}
\end{align}

The anti-symmetric tensors in three dimensions are spanned by the Hodge 
duals of the vectors $\by$, $\bt$ and $\bt\times\by$. That is, they are 
spanned by  
\begin{align}
A^{ij} &:= t^i y^j- t^j y^i, \quad B^{ij} := \eps^{ijk} y_k, \word{and} 
C^{ij} := \eps^{ijk} t_k, \nn \\ 
\word{or} A&= \bt\wedge\by = \ast(\bt\times \by), 
\quad B= \ast \by, \word{and} C=\ast \bt.
\label{eq:asymtensor}
\end{align}
For convenience, note that the tensors 
\begin{align}
D^{ij} &:= A_{kl}( \eps^{ikl} y^j - \eps^{jkl} y^i)
\word{and}  F^{ij} := A_{kl}( \eps^{ikl} t^j - \eps^{jkl} t^i)
\end{align}
satisfy 
\begin{align}
\frac{1}{2} D^{ij} &= {y^0} t^0 B^{ij} - y^2 C^{ij}; 
\nn \\
\frac{1}{2} F^{ij} &= \frac{{y^0}^2}{r^2R^2} B^{ij} - y^0 t^0 C^{ij}.
\end{align}

\begin{remark}
In the case of a $k=-1$ spacetime, where we look for antisymmetric rank 
two tensors in four spacetime dimensions, invariant under $SO(3,1)$, we 
only have two independent candidates $A_4= t\wedge y$ and $B_4 = \ast(t\wedge y)$, 
while in the case of $k=0$, three spatial dimensions and $SO(3)$, 
we have the three candidates \eqref{eq:asymtensor}.
\end{remark}

\paragraph{Minimal case \texorpdfstring{\boldmath{$n=0$}}{}.}
Starting from the form \eqref{eq:mmunu_minimal}, the formula \eqref{eq:m_gens} 
was proven in \cite{Gass:2025tae} for the minimal case $n=0$.

\paragraph{Large \texorpdfstring{\boldmath{$n$}}{} case.}
Substituting $y^\mu$ and $t^\mu$ for $x^\mu$ and $t_{k=-1}^\mu$ into 
\eqref{eq:mnunu-1} yields\footnote{Note that one needs to be careful raising and 
lowering indices: The indices of $y^\mu$ and $x^\mu$ are lowered with a different 
Minkowski metric. In the following, all indices are raised and lowered with the 
Minkowski metric corresponding to $x^\mu$. This does not matter for the spatial 
components, but we make sure that $x^0$, $t^0$ and $y^0$ do only appear with 
upper indices in this derivation.}
\begin{align}
 m^{ij} 
 &= \frac{Rx^4}{R^2+{x^4}^2} A^{ij} -\frac{R^2 t^0}{R^2+{x^4}^2} B^{ij}
    + \frac{ R^2x^0}{R^2+{x^4}^2} C^{ij} \nn \\
   &\quad  + \frac{x^4}{R^2+{x^4}^2} (y^i m^{j0} - y^j m^{i0})
           -  \frac{R x^0}{R^2+{x^4}^2} \eps^{ij}_{\phantom{ij}k}
           m^{k0}.
\end{align}
A similar substitution yields 
\begin{align}
m^{i0} 
 &= \frac{Rx^4}{R^2 + {x^4}^2} \big(x^0 t^i - \tfrac{1}{R}x^0 m^{i0} 
        - t^0 y^i\big) 
          + \frac{R^2}{R^2 + {x^4}^2} \eps^{ijk} y_j t_{k} 
          - \frac{R
          }{R^2 + {x^4}^2} \eps^{ij}_{\phantom{ij}k} y_j m^{k0} \nn \\ 
&= \frac{1}{R^2 + x^4y^0} 
\Big( R x^4 (x^0 t^i - t^0 y^i ) 
 + R^2 \eps^{ijk} y_j t_k - R \delta^{il} \eps_{ljk} y^j m^{k0} \Big).
\end{align}
Now, the vector $V^i := m^{i0}$ must be a linear combination 
\begin{align}
m^{i0} = a y^i +b t^i + c \eps^{ijk} y_j t_k,
\end{align}
where $a,b,c$ are functions of $y^0,t^0,y$. Inserting this form 
into the first form of $m^{i0}$ and equating the two yields 
\begin{align}
&\quad
- Rt^0 ( x^4  + c y^0) y^i 
    + R ( c y^2 + x^4 x^0) t^i
    + R(R-b) \eps^{ijk} y_j t_k \nn \\
&= (R^2 + x^4y^0) (ay^i + b t^i + c \eps^{ijk} y_j t_k).
\end{align}
The corresponding system of equations is uniquely solved by 
\begin{align}
a = - \frac{Rt^0}{y^0},\quad 
b  = R\frac{{y^0}^2+y^2-R^2}{2{y^0}^2}, \word{and}
c  = \frac{R^2}{{y^0}^2},
\end{align}
which implies \eqref{eq:m_gens}.

Using \eqref{eq:m_gens}, one then easily verifies $\{m^{ij},y^0\}
=\{m^{ij},t^0\}=0$ and $\{m^{ij},t^k\} = \delta^{ik} t^j - \delta^{jk} t^i$.
Finally, we have 
\begin{align}
\{m^{ij},y^k\} 
&= \delta^{ik} y^j -  \delta^{jk} y^i 
+ \frac{r^2R^3}{{y^0}^2} \Big(  
  \eps^{ikl}  t^j  -  \eps^{jkl}  t^i  - \eps^{ijl}  t^k \Big)  t_l
 +  R \eps^{ijk} \nn \\  
 &= \delta^{ik} y^j -  \delta^{jk} y^i,
\end{align}
which implies $\{m^{ij},y\}=0$.

\paragraph{Acknowledgments.}
This work is supported by the Austrian Science Fund (FWF) grant P36479. HS would 
like to thank Pei-Ming Ho and Hikaru Kawai for a related collaboration. 
We also thank Alessandro Manta for valuable discussions.

\footnotesize
\bibliography{bibliography}{}
\bibliographystyle{unsrt}

\end{document}